\documentclass[12pt]{article}
\usepackage{amsmath,epsf,amssymb,latexsym,cite,xy}
\xyoption{matrix}\xyoption{arrow}

\usepackage[dvips]{epsfig}

\newcommand{\beq}{\begin{equation}}
\newcommand{\eeq}{\end{equation}}
\newcommand{\beqa}{\begin{eqnarray}}
\newcommand{\eeqa}{\end{eqnarray}}

\newcommand{\cN}{\ensuremath{\mathcal{N}}}

\newif\iffigs\figstrue

\DeclareFontFamily{U}{rsf}{}
\DeclareFontShape{U}{rsf}{m}{n}{
   <5> <6> rsfs5 <7> <8> <9> rsfs7 <10-> rsfs10}{}
\DeclareMathAlphabet\Scr{U}{rsf}{m}{n}

\def\pplogo{\vbox{\kern-\headheight\kern -43pt
\halign{##&##\hfil\cr&{%\sc
\ppnumber}\cr\rule{0pt}{2.5ex}&\ppdate\cr}
}}
\makeatletter
\def\ps@firstpage{\ps@empty \def\@oddhead{\hss\pplogo}%
   \let\@evenhead\@oddhead % in case an article starts on a left-hand  page
}
\def\maketitle{\par
  \begingroup
  \def\thefootnote{\fnsymbol{footnote}}
  \def\@makefnmark{\hbox{$^{\@thefnmark}$\hss}}
  \if@twocolumn
  \twocolumn[\@maketitle]
  \else \newpage
  \global\@topnum\z@ \@maketitle \fi\thispagestyle{firstpage}\@thanks
  \endgroup
  \setcounter{footnote}{0}
  \let\maketitle\relax
  \let\@maketitle\relax
  \gdef\@thanks{}\gdef\@author{}\gdef\@title{}\let\thanks\relax}
\makeatother

\def\O{\Scr{O}}

\def\R{{\mathbb R}}

\def\tr{\operatorname{tr}}

\def\cD{{\Scr D}}

\def\cH{{\Scr H}}

\def\cE{{\Scr E}}

\def\cH{{\Scr H}}

\def\cf{{\em c.f.}}

\def\LARGE{\large\bf}

\def\CN{{\cal N}}

\def\ket#1{{| #1 \rangle}}
\def\vev#1{{\langle #1 \rangle}}
\def\p{\partial}
\def\sig#1{\sigma_{#1}}
\def\half{\frac {1}{2}}
\def\del{\nabla}

\def\eg{{\it e.g.}}

\begin{document}

\setcounter{page}0
\def\ppnumber{\vbox{\baselineskip14pt
\hbox{BRX-TH-556}
\hbox{hep-th/0508148}}}
\def\ppdate{August 2005} \date{}

\title{\LARGE Random walks and the Hagedorn transition\\[10mm]}
\author{
Martin Kruczenski${}^{1,2}$ and Albion Lawrence${}^1$ \\[2mm]
\normalsize ${}^1$ Martin Fisher School of Physics, Brandeis University, \\
\normalsize MS057, PO Box 549110, Waltham, MA 02454\\[2mm]
\normalsize ${}^2$ Dept. of Physics, Princeton University\\
\normalsize Princeton, NJ 08544\\
\normalsize ({{\em Present address}})
}

{\hfuzz=10cm\maketitle}

\begin{abstract}
We study details of the approach to the Hagedorn temperature 
in string theory in various static  spacetime backgrounds.  
We show that the partition function for a {\it single} string at finite temperature  
is the torus amplitude restricted to unit winding around Euclidean time.   
We use worldsheet path integral to derive the statement that the
the sum over random walks of the thermal scalar near the Hagedorn transition
is precisely the image under a modular transformation of the sum over spatial configurations 
of a single highly excited string. We 
compute the radius of gyration of thermally excited strings in  $AdS_D\times S^n$.   
We show that the winding mode
indicates an instability despite the AdS curvature at large radius, and that
the negative mass squared decreases with decreasing AdS radius, much like
the type 0 tachyon.  We add further arguments to statements by 
Barb\'on and Rabinovici, and by Adams {\it et. al.}, that the Euclidean AdS black hole can
thought of as a condensate of the thermal scalar.  We use this to provide circumstantial
evidence that the condensation of the thermal scalar decouples closed string modes.
\end{abstract}

\vfil\break

%\maketitle

\section{Introduction}

Polchinski \cite{Polchinski:1985zf} has shown that 
the free energy of a gas of noninteracting strings on a spatial manifold $M$
at temperature $T=\beta^{-1}$ is
the partition function of a single string on a torus, in the Euclidean background 
$M\times  S^1_{\beta}$.  Here the circle
$S^1_{\beta}$ is the Euclidean time direction and has radius $\beta$.  As  
the temperature increases from below, 
there is a scalar string state known as the "thermal scalar"
with unit winding  around  $S^1_{\beta}$ which becomes a spacetime tachyon
\cite{Sathiapalan:1986db, Kogan:1987jd,Atick:1988si} 
at the "Hagedorn temperature" $T_H \sim m_s = \ell_s^{-1}$. (Here $\ell_s$ is the string scale
and $m_s^2$ is the string tension).
Near the Hagedorn temperature, this scalar dominates the partition function \cite{Atick:1988si,Brandenberger:1988aj,Horowitz:1997jc}.
If there are noncompact spatial dimensions, then the divergence of the partition function 
near a phase transition arises from the infrared divergence of the 
one-loop contribution of the tachyon. Even if all dimensions are compactified, the
one-loop partition function for the string gas diverges as $\ln(T_H - T)$ due to the appearance of the tachyon.

On the other hand, in the presence of noncompact spatial dimensions,
an ensemble of strings at fixed energy is dominated by configurations  
with a {\it single} long string and a gas of small, light strings
\cite{Mitchell:1987hr,Mitchell:1987th,Bowick:1989us}.  
The radius of gyration of a long string with energy $E$ and tension  
$\ell_s^{-2}$ is $\vev{r^2}\sim \ell_s^3 E$, indicating that the statistics of these strings
are those of random walks of length $L = \ell_s^2 E$.

The thermal scalar seems like a formal device, and it would be nice  
to better understand the relation between the canonical and microcanonical  
descriptions of a gas of strings.  One hint appears in \cite{Horowitz:1997jc,Barbon:2004dd}: if one  
computes the one-loop partition function of the thermal scalar near the phase transition, it can be  
rewritten as the partition function of a single string with density of states
\begin{equation}
    \rho(E) = \frac{e^{\beta_H E}}{E^{1 + \alpha}}\ ,
\end{equation}
where $\alpha$ depends on the details of the worldsheet conformal field  
theory, such as the number of noncompact dimensions.\footnote{For the purposes
of this paper, "noncompact dimensions"
is a code for the presence of a continuous spectrum of conformal dimensions
on the worldsheet, or of a continuous spectrum for the momentum
operator.  From this perspective, for example, anti-de Sitter space should be
considered "compact".}  Furthermore, the correlation  
function of thermal scalars can be computed via a sum over random walks.

Atick and Witten \cite{Atick:1988si}\ argued, by analogy with the deconfinement transition in QCD,
that this transition should give some insight into the fundamental degrees of freedom
of string theory.  Since equilibrium thermodynamics
in flat space fails in the presence of gravity, it is hard to know how
far one can go with the perturbative flat space calculations.  Anti-de Sitter (AdS) space,
however, provides a convenient "box" for gravity, and the anti-de Sitter/conformal field
theory (AdS/CFT) correspondence \cite{Maldacena:1997re}\
gives a definite prescription for studying string theory at finite temperature in such backgrounds.
In light of this correspondence, one finds that there is indeed a relation between the Hagedorn
transition and deconfinement
\cite{Sundborg:1999ue,Haggi-Mani:2000ru, Aharony:2003sx,Spradlin:2004pp,Liu:2004vy,
Aharony:2005bq, Alvarez-Gaume:2005fv, Schnitzer:2004qt, Gomez-Reino:2005bq}, 
which we will explore further below.  

The goal of our paper is twofold.  First, we will make the relation between
the thermal scalar and random walks explicit, 
expanding on \cite{Horowitz:1997jc,Barbon:2004dd}, and apply our lessons
to strings in AdS backgrounds.  Next, we will tie together
some of our own calculations for string theory in finite-temperature AdS backgrounds
with previous work by other authors to flesh a picture of the Hagedorn
transition in AdS backgrounds.

More precisely, in \S2 we show that the partition function 
for a single string at finite temperature can be written as  
the worldsheet torus amplitude restricted to unit winding around the torus.  If
there are "noncompact" directions in the target space -- directions
in which the spectrum of the momentum operator is continuous -- then the 
the single-string dominance of the microcanonical ensemble and the leading infrared
singularity of the  thermal scalar contribution 
near the Hagedorn temperature are the same effect, related by
a worldsheet modular transformation.  In completely compactified spacetimes, or
spacetimes like $AdS_D\times S^n$ for which the momentum operator is discrete,
single string dominance fails to hold in the microcanonical ensemble, while
in the canonical ensemble the large-volume divergence of the thermal scalar
is stripped away leaving the divergence due to the existence of an unstable mode.
In  \S3\ we explicitly derive, in a wide class of static spacetimes, the statement that the 
sum over the spatial configurations of highly energetic strings is precisely a 
sum over random walks.
We compute the size of highly excited strings in $D$-dimensional
anti-de Sitter space ($AdS_D$) times an $n$-sphere $S^n$, 
as a function of temperature; and 
we show that the tachyon indicates a genuine instability in AdS spacetimes.
In \S4\ we extend arguments of \cite{Barbon:2001di,Barbon:2002nw,Barbon:2004dd,Adams:2005rb}
that the endpoint of the condensation of the tachyon is the AdS black hole.  We then
provide some circumstantial evidence that the closed string modes decouple
in the presence of the tachyon in much the same way that open string
excitations decouple when the open string tachyon condenses
\cite{Sen:1998sm,Sen:1999md,Sen:1999xm,Sen:1999nx,Berkovits:2000hf,Yi:1999hd,
Bergman:2000xf,Gopakumar:2000rw,David:2000uv,Taylor:2000ku,Sen:2000hx,Kleban:2000pf}.
In \S5 we give a brief conclusion. 

We should note that many of the basic statements in \S2,\S3\ are contained implicitly or explicitly in
\cite{Horowitz:1997jc,Barbon:2004dd}; while many of the statements in \S4.2\ are contained
in recent work such as \cite{Barbon:2001di,Barbon:2002nw,Barbon:2004dd,Adams:2005rb, 
Horowitz:2005vp, McGreevy:2005ci}, and are surely known to others. However, we feel that our explicit derivation, the expanded
discussion, and consolidation are worth putting into print. Other recent work on the relation between excited
strings and random walks is \cite{Manes:2004nd}.

%\section{Review of strings at finite temperature}

\section{The partition function for single strings}

Polchinski showed that the partition function for a gas of strings
in the canonical ensemble at temperature $T = \beta^{-1}$ can be computed
as follows.  Let the worldsheet conformal field theory consist of
a $c=25$ (for the bosonic string) or $\hat{c}=9$ (for the superstring)
factor for the spatial directions of the target space, times a $c=1$ or $\hat{c} = 1$
CFT for the target  
space circle $S^1_{\beta}$ with radius $\beta$.  The free energy
$\beta F(\beta)$ is the vacuum torus amplitude for the string in this Euclidean background  
\cite{Polchinski:1985zf}.  One indication of this is that the amplitude can be written as the sum  
over closed  string modes of the free energy for a gas of particles in each mode.

We will argue that the partition function for a  
{\it single} string at temperature $\beta$ is the vacuum amplitude in Euclidean space
with periodic Euclidean time, restricted to unit winding
around $S^1_{\beta}$.

\subsection{Partition function for a single string}

Consider a scalar particle, or a particle in a fixed eigenstate of
spin angular momentum, with mass $m$ in flat
space. The Euclidean spacetime is $M_{d+1} = S^1_{\beta} \times  
\mathbf{R}^d$.
The partition function for one particle (as opposed to a gas of  
particles) with mass $m$ in $d$ spacetime dimensions is:
\begin{equation}\label{eq:onepartpart}
    Z = \int \frac{d^{d-1}k}{(2\pi)^{d-1}}
     e^{-\beta \omega_k}\ \ \ \ \ \ \omega_k^2 = \vec{k}^2 + m^2
\end{equation}
This can be written as a space-time integral\footnote{This relation can  
be derived in various ways. An alternative
is to use the identity
$e^{- 
\beta\sqrt{m^2+k^2}}=\frac{\beta}{\sqrt{2\pi}}\int_0^{\infty}\frac{ds}{s 
^{3/2}}e^{-\half(m^2+k^2)s}e^{-\frac{\beta^2}{2s}}$}:
\begin{eqnarray}\label{eq:oneparticlepf}
    Z &=& \int \frac{d^dk}{(2\pi)^d} \frac{2ik^0 e^{i\beta k^0}}{k^2 +  
m^2} \ ; \ \ \ \ \ \ \ \ 
    (k^2 = (k^0)^2 + \vec{k}^2 + m^2) \cr
    &=& \int \frac{d^d k}{(2\pi)^d} \int^{\infty}_{0} ds
    (2ik^0) e^{i\beta k^0 - s((k^0)^2 + \vec{k}^2 + m^2)}\cr
    &=& 2\p_{\beta} \int_0^{\infty} \frac{ds}{(2\pi s)^{d/2}}  
e^{-\frac{\beta^2}{4s} - m^2 s}\cr
    &=& - \beta \int_0^{\infty} \frac{ds}{s (2\pi s)^{d/2}}  
e^{-\frac{\beta^2}{4s} - m^2 s}
\end{eqnarray}
For a single string, one can sum this answer over all possible  
string states.  The only difference between this  
calculation and that of \cite{Polchinski:1985zf}\ is that we are looking at  
worldlines that wrap {\it once} around the Euclidean time direction.  Therefore, we can  
write the partition function for a single string at spacetime temperature $T = \beta^{-1}$ as:
\begin{equation}\label{eq:onepmodular}
    Z_{s.s.} = \int_0^{\infty} \frac{d\tau}{4\pi \tau_2^2}
        \int_{-\half}^{\half} d\tau_1 I_{(0,1)}(\tau)
\end{equation}
where $I_{(m,n)}(\tau)$ is the one-loop partition function for the conformal field theory
on the torus with modular parameter $\tau = \tau_1 + i \tau_2$, coordinates
$\sigma_1 \in [0,1], \sigma_2\in[0,1]$, worldsheet metric
$ds^2 = |d\sigma_1 + \tau d\sigma_2|^2$, and boundary conditions
\begin{eqnarray}
    X^{\mu} (\sig{1} + 1, \sig{2}) &=& X^{\mu}(\sig{1},\sig{2}) +  
m\beta\delta^{\mu,0}\cr
    X^{\mu} (\sig{1},\sig{2}+1) &=& X^{\mu}(\sig{1},\sig{2})
        + n\beta\delta^{\mu,0}\ .
\end{eqnarray}

Note that the integral is not over the fundamental domain of $\tau$ but  
rather along the entire strip $\tau_2 \geq 0$, $-\half \leq \tau_1 \leq \half$.  
This is the fundamental domain of the subgroup $\tau \to \tau + 1$ of the full
modular group.  Under this transformation (\cf\ \cite{McClain:1986id}) the
winding numbers $(m,n)$ are mapped to $(m,n-m)$.  Therefore the $(0,1)$ sector
is invariant under this shift and one should restrict the modular integral over $(0,1)$ to
an integral over this strip.

One may further include the effects of the modular transformation $\tau \to 1/\tau$.
Under thi-s transformation, however, $(m,n)$ is mapped to $(n,-m)$; in particular
$(0,1)$ is mapped to $(1,0)$.  The $(0,1)$ sector is not invariant under the
full modular group $SL(2,\mathbb{Z})$.  Its image is the set
of {\it coprime}\ integers $(m,n)$.  One may therefore rewrite the above partition
function as an integral over the
fundamental domain ${\cal{F}} = \{\tau \big| |\tau_1| \leq  
\half, |\tau|^2 \geq 1\}$, and a sum over $I_{m,n}$ with $(m,n)$ coprime:
\begin{equation}\label{eq:singlestringpf}
    Z_{s.s.} = -\beta \int_{\cal{F}} \frac{d^2\tau}{4\pi \tau_2^2}
        \sum_{(m,n)\ coprime} I_{(m,n)}(\tau)\ .
\end{equation}

\subsection{Single string dominance and the Hagedorn transition}

If we sum (\ref{eq:singlestringpf}) over all $(m,n)\in \mathbb{Z}^2$, the result
is $-\beta$ times the free energy for a {\it gas} of strings at  
temperature $\beta^{-1}$ \cite{Polchinski:1985zf}.  If $Z_k$ is the
partition function for $k$ strings, and we fix our normalization so that
$Z_0 = 1$, then the free energy can be written as
\begin{eqnarray}\label{eq:freetopart}
    F & = & - \frac{1}{\beta} \ln \left(1 + Z_1 + Z_2 + \ldots\right) \cr
    &=& - \frac{1}{\beta} \left(Z_1 + (- \half Z_1^2 + Z_2) + \ldots  \right)\cr
    &=& - \frac{1}{\beta} \left(Z_{[0,1]} + Z_{[0,2]} + \ldots\right)\ .
\end{eqnarray}
Here $Z_{[0,k]}$ comes from replacing $I_{(0,1)}$ in (\ref{eq:onepmodular})
with $I_{(0,k)}$, and $Z_{[0,1]} = Z_{s.s.}$.  We will demonstrate in Appendix
A that $Z_2 - \half Z_1^2 = Z_{[0,2]}$.  $Z_{[0,k]}$ looks effectively like the partition
function for a single string at temperature $1/(k\beta)$.  Therefore, $Z_{[0,1]} = Z_{s.s.}$
dominates the free energy for a gas of strings at the Hagedorn temperature.
We should note that the partition function
$Z = \sum_k Z_k$ generally contains divergences from each sector $Z_k$,
although when there are noncompact spatial dimensions
(\eg\ the spectrum of conformal dimensions is continuous) 
the divergences will be subleading for $k>1$: it is well known that
single strings dominate the partition function for very high energies
\cite{Mitchell:1987hr,Mitchell:1987th,Bowick:1989us}, and
it is the high energy states which dominate near the Hagedorn transition.

Strings in the sector $(m,n) = (1,0)$ become tachyonic at the
"Hagedorn temperature" $(\beta_H)^{-1} = (2\sqrt{2}\pi \ell_s)^{-1}$.
If the number of noncompact spatial
dimensions is nonzero, this leads to an infrared divergence in the partition function  
characteristic of a massless scalar field, characteristic of a phase transition \cite{Atick:1988si}.
This divergence is related under $\tau\to \frac{-1}{\tau}$
to the ultraviolet divergence coming from the $(m,n) = (0,1)$ sector;
this divergence arises from the large UV density of states of
a single string.  In other words, the divergence in $Z_1$ is mapped
by a modular transformation to the infrared divergence in the free energy for
the thermal scalar. We will explore this map in detail in \S3.

Furthermore, the large-volume infrared divergence for the thermal tachyon 
is cut off when all directions are
compactified.  It is known that in this case, a single long  
string no longer dominates the ensemble at large energies 
\cite{Bowick:1989us, Barbon:2004dd}. One is left with a divergence
due entirely to the negative eigenvalue of the tachyon kinetic term.

In short, in the presence of noncompact dimensions,
the infrared divergence in $F$ due to the thermal scalar
at the Hagedorn temperature is a re-encoding of 
single-string dominance, via a worldsheet modular transform.

\section{The random walk picture for excited strings}

\subsection{Review of the random walk picture of the thermal scalar}

Studies of a gas of strings in the microcanonical ensemble  
\cite{Mitchell:1987hr,Mitchell:1987th,
Bowick:1989us} have shown that the single string which dominates at  
large energies has a radius of gyration $r_g$ characteristic of a random  
walk, $r_g^2 = \vev{\delta x^2} \sim \ell_s L = \ell_s^3 E$, where
$L$ is the length of the string, $E$ the energy, and $m_s^2 = \ell_s^{-2}$ the string 
tension.  Horowitz and Polchinski \cite{Horowitz:1997jc}\ have also argued this from the  
canonical ensemble. The essence of their 
argument can be summarized as follows.  The quadratic part of the
thermal scalar Lagrangian is
$-\half \int d^d x \phi D \phi$, with $D = \del^2 + m_{\beta}^2$ and
$m_{\beta}^2 = \frac{\beta_H^2 - \beta^2}{4\pi(\alpha')^2}$  
\cite{Horowitz:1997jc}.  The equation for the heat kernel of D:
\begin{equation}
    (\del^2  + m_{\beta}^2) K = \ell_s \partial_T K; \ \ \ K(\vec{x},T = 0)  
= \delta^{(d)}(\vec{x})
\end{equation}
can be solved by writing $K = e^{m^2 \ell_s T} P$. Now,
\begin{equation}
    P = \frac{e^{-\frac{|x|^2}{2\ell_s T}}}{(\ell_s T)^{d/2}}
\end{equation}
solves the diffusion equation in $d$ dimensions,
which is also the equation for the probability distribution of the  
position of a random walk of length $T$ which begins at the origin.  Near  
$m^2_{\beta} = 0$, we can write
\begin{equation}
    m_{\beta}^2 = \frac{\beta_H^2 - \beta^2}{4\pi^2(\alpha')^2}
        \sim \frac{\beta_H(\beta_H - \beta)}{2\pi^2(\alpha')^2}
\end{equation}
Therefore, we find that
\begin{equation}
    K = e^{\frac{\beta_H^2}{2\pi(\alpha')^2}\ell_s T}P(\vec{x},T)
    e^{-\frac{\beta_H \ell_s T}{2\pi(\alpha')^2} \beta}
\end{equation}
Here $\beta_H = \alpha'$, the first factor $e^{\beta_H^2 \ell_s T/2\pi(\alpha')^2}$ on the right hand side
is the total  number of random walks of length $T$, and  the third factor
$e^{-\beta_H \beta\ell_s T / 2\pi(\alpha')^2}$ is a  
Boltzmann suppression factor.

The free energy for the thermal scalar is
\begin{equation}\label{eq:heatkern}
    F = -\beta \ln Z = -\beta \int_0^{\infty} \frac{dT}{T} K(0,T)
\end{equation}
Here $e^{\beta_H^2\ell_s T/2\pi(\alpha')^2}P(0,T)/T$ is the number of  
closed paths of length $T$: the extra factor of $1/T$ compensates for the fact  
that one may choose any point on the loop as the starting place for the walk.   
The result is that the free energy for the thermal scalar is precisely the  
partition function for a single static string with tension $1/2\pi(\alpha')$.

We would like to relate this result to the sum over massive string states in the
following way.  As discussed in \S2, the worldsheet path integral on
the torus with winding number $(1,0)$ contains the path integral
over the thermal scalar.  Close to the Hagedorn temperature,
the path integral is dominated by large values of $\tau_2$.  We
identify this with the Schwinger parameter $T$ in the
discussion above. However, as we noted in \S2,  the 
modular transformation $\tau \to - \frac{1}{\tau}$ takes this path integral
to one over a string with winding number $(0,1)$.  Large $\tau_2$ is mapped
to small $\tau_2$.  The sum over paths of the thermal scalar are mapped
to sums over spatial configurations of the string, which wraps the thermal
circle once as in Fig. 1.  The random walks summed over in the thermal
scalar path integral should be precisely the spatial configurations of the string
in the first-quantized picture.

\begin{figure}
\centering
\epsfig{file=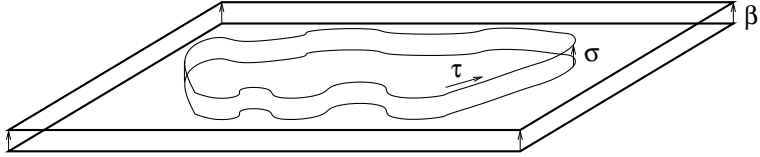, width=10cm}
\caption{{\it The path integral for a single string close to the
Hagedorn temperature can be written as sums over random walks
of the thermal winding mode, or as random spatial configurations
of the string which wraps once around the thermal time direction.}}
\label{fig:largeT} 
\end{figure}

We would like to understand the degree to which this picture is precise.  In the string theory
path integral, one can write worldsheet coordinates $t,\sigma$ such that
the winding of the string is in the $t$ direction.  A simple sum over random walks would
be a sum over string worldsheets for which the target space coordinates of the string
depend only on $\sigma$: however, in general the coordinates can depend both on the spatial 
coordinate $\sigma$ on the worldsheet, and on the time direction $t$, as in Fig. 2. 
Nonetheless, for generic highly massive strings we will directly derive 
this picture in a way that will generalize to 
strings in a large class of static curved spacetimes, such as anti-de Sitter space,
so we can compute the radius of gyration of a string as a function of temperature in 
these spacetimes as well.
Our general results will be essentially those stated in \cite{Horowitz:1997jc},
though we hope that our explicit derivation and discussion of its validity will
be of use.   We will go on to derive the radius of gyration of strings in anti-de Sitter space
at finite temperature.

\begin{figure}
\centering
\epsfig{file=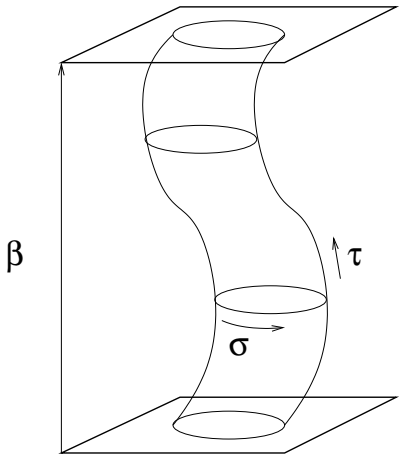, height=6cm}
\caption{{\it Periodic worldsheet contributing to the finite temperature path integral (the upper and lower 
surfaces are identified). At temperatures away from the Hagedorn temperature,
the string oscillations will depend on Euclidean time.}}
\label{fig:smallT} 
\end{figure}

\subsection{Path integral for highly excited strings in curved spacetimes}

We study string theory on the target space $S^1_{\beta} \times M_9$, parameterized by
coordinates $X^{0,\ldots, d-1}$, where $X^0 \equiv X^0 + 2\pi\beta$.  We will take
target space metrics of the form:
\begin{equation}
    ds^2 = G_{\mu\nu}dX^{\mu}dX^{\nu} = G_{00}(X^i) (dX^0)^2 + G_{ij}(X^i)\,dX^i dX^j
\label{eq:staticmetric}
\end{equation}
In particular we assume that translations in $X_0$ are isometries of the metric.
We are interested in configurations such that $X^{\mu}(\sigma^1+1,\sigma^2) = X^{\mu}$,
$X^0(\sigma^1,\sigma^2+1) = \beta + X^0(\sigma^1,\sigma^2)$, and
$X^i(\sigma^1,\sigma^2+1) = X^i(\sigma^1,\sigma^2)$.
The Polyakov action for strings on the torus in conformal gauge is:
\begin{equation}\label{eq:torusactionone}
	S = m_s^2 \int d^2\sigma \left[ \frac{|\tau|^2}{\tau_2}
		G_{\mu\nu}\p_1X^{\mu}\p_1X^{\nu}
		- 2 \frac{\tau_1}{\tau_2} G_{\mu\nu}\p_1X^{\mu}\p_2X^{\nu}
		+ \frac{1}{\tau_2}G_{\mu\nu}\p_2X^{\mu}\p_2X^{\nu}\right]\nonumber \\ 
\end{equation}
We wish to study the full partition function
figure\begin{equation}
	Z = \int_{-\frac{1}{2}}^{\frac{1}{2}} d\tau_1 \int_0^{\infty}
	\frac{d\tau_2}{\tau_2^2} \Delta_{FP} \int \cD X \prod_{\sigma^1,\sigma^2}
	\sqrt{G(X(\sigma^{\alpha}))} e^{-S[X]}
\label{partfunct}
\end{equation}
in the small $\tau_2$ region of the integrand, which dominates for highly excited string states.
Note that we have kept the $\sqrt{G}$ measure factor explicit, so that the
path integral measure is invariant under reparameterizations of the target space.
The Fadeev-Popov determinant $\Delta_{FP}$ is given by
\beq
\Delta_{FP} = \frac{1}{\tau_2} {\det}' \Delta
\eeq
with $\Delta$ the worldsheet Laplacian. The prime in the determinant indicates that we have
removed the zero mode.

\begin{figure}
\centering
\epsfig{file=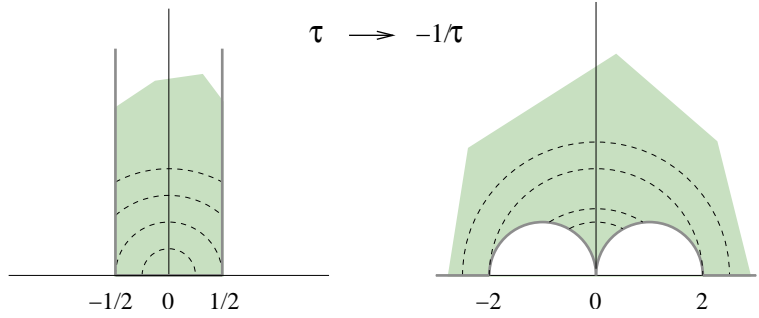, width=10cm}
\caption{{\it The domain $|\Im m \tau| \leq \half$ transforms under $\tau\to -1/\tau$
as shown.  A simple way to undersand the mapping is that circles around the origin map to circles
around the origin but with inverse radius (see the dashed circles in the figure).}}
\label{fig:modularflip} 
\end{figure}

We will make a change of variables $\tau\to - \frac{1}{\tau}$ in the integrand, 
which is of course just a modular transformation.  Note that the $\tau$ 
measure $\frac{d^2\tau}{\tau_2^2}$, the Faddev-Popov determinant
$\Delta_{FP}$, and the path integral over $X$ are separately modular invariant. The integration 
domain is mapped to that shown in Fig. \ref{fig:modularflip}.  The boundary conditions on $X^{\mu}$ become:
$X^{\mu}(\sigma^1,\sigma^2+1) = X^{\mu}$,
$X^0(\sigma^1+1,\sigma^2) = \beta + X^0(\sigma^1,\sigma^2)$, and
$X^i(\sigma^1+1,\sigma^2) = X^i(\sigma^1,\sigma^2)$, and we will
set $X^0 = \beta\sigma^1 + \delta X^0$, where $\delta X^0$ is periodic on the torus.
The resulting action on the torus is:
\begin{eqnarray}
	S &=& m_s^2 \int d^2\sigma \left[ \frac{|\tau|^2}{\tau_2}
	\left(G_{00}\left(\beta^2 + 2\beta \p_1 X^0 +(\p_1 X^0)^2\right) + 
	G_{ij}\p_1X^i\p_1 X^j\right)\right. \nonumber\\
	& &\ \ \ \ -2 \frac{\tau_1}{\tau_2}\left(G_{00} \beta \p_2 \delta X^0 + 
		G_{00}\p_1\delta X^0\p_2\delta X^0 +
		G_{ij}\p_1X^i\p_2X^j\right) \nonumber \\
	& & \ \ \ \ \left. + \frac{1}{\tau_2}\left( G_{00}(\p_2\delta X_0)^2 +
		G_{ij} \p_2 X^i \p_2 X^j \right)\right] \label{eq:torusactiontwo}
\end{eqnarray}

Having made a the transformation $\tau \to -1/\tau$,
we are interested in the limit of {\em large} $\tau_2$. In that limit 
from eq.(\ref{eq:torusactiontwo}) we see that configurations such that $\p_1 \delta X^0\neq 0$ and/or 
$\p_1 X^i\neq 0$ are highly supressed in the path integral.  However, one cannot completely
ignore the effects of these terms: in the limit of large $\tau_2$, the oscillator modes
contribute to the vacuum energy and thus lead to a contribution
$e^{c\pi/6}$ to the path integral, where $c$ is the number of oscillators.
%\footnote{This
%is closely related to the high-temperature limit of quantum statistical mechanics for
%$n+1$-dimensional theories -- \cf\ 
%\cite{Atick:1988si,Yaffe:1982qf}.  Naively this limit is described by dimensional
%reduction on the thermal circle: however, the modes with momentum along this circle contribute
%to the renormalization of some of the quantities of the $n$-dimensional theory.}
We can combine this factor with the large-$\tau$ limit of $\Delta_{FP}$:
\beq
 \frac{1}{\tau_2} {\det}' \Delta\ {\stackrel{\tau_2 \to \infty}{\sim} }\ \tau_2 e^{-\frac{\pi}{3}\tau_2} 
\eeq 
For the case of the superstring, 
the superconformal ghosts and fermion superpartners will lead to similar terms in the path integral.
Their effect is to add a term of the form $\delta S = - \int_0^1 d\sigma^2 m_s^2 \beta_H^2 \tau_2$
to the action.  If we include this term and set all the $\p_1$ derivatives
to zero, we can write the resulting action as:
\begin{equation}
	S = m_s^2 \int d\sigma_2 \left[ \frac{1}{\tau_2} (G_{ij}\p_2 X^i \p_2 X^j)
	+ \tau_2 \left( G_{00} \beta^2 - \beta_H^2\right)+ 
	\frac{\beta^2 G_{00}}{\tau_2} \left(\tau_1 - \frac{\p_2 	
		X^0}{\beta}\right)^2\right]
\label{resaction}
\end{equation}
The path integral over $X^0$ is Gaussian and can be evaluated exactly. The equation of motion 
for $X^0$ is 
\beq
\p_2 \left[  G_{00}  \left(\tau_1 - \frac{\p_2 	
		X^0}{\beta}\right)\right] = 0
\eeq
with solution
\beq
\p_2 X^0 = \beta \tau_1 - \frac{C}{G_{00}}
\eeq
where $C$ is a contant of integration which can be found using the condition that $X^0$ is periodic in $\sigma_2$:
\beq
0= \int_0^1 d\sigma_2 \, \p_2 X^0  \ \ \ \ \Rightarrow \ \ \ \ C = \beta \tau_1 
\left(\int_0^1\frac{d\sigma_2}{G_{00}}\right)^{-1}
 \equiv \beta\tau_1\langle G_{00}^{-1}\rangle^{-1}\ ,
\eeq
where we define $\langle F\rangle = \int_0^1 d\sigma_2 F$.
After we expand around this classical solution, the path integral over $X_0$ gives
\beqa
\int \cD X^0(\sigma_2) e^{-\frac{m_s^2}{\tau_2}\int d\sigma_2 G_{00} (\beta\tau_1-\p_2X^0)^2} &=& 
e^{-\frac{m_s^2\beta^2\tau_1^2}{\tau_2} \langle G_{00}^{-1}\rangle^{-1}} 
 \int \cD X^0(\sigma_2) e^{-\frac{m_s^2}{\tau_2}\int d\sigma_2 G_{00} (\p_2X^0)^2} \\
 & =& \cN_0 \tau_2^{-\half}\langle G_{00}^{-1}\rangle^{-\half}  
         e^{-\frac{m_s^2 \beta^2\tau_1^2}{\tau_2} \langle G_{00}^{-1}\rangle^{-1}}   
\label{eq:timeintegral}
\eeqa
where $\cN_0$ is an infinite normalization factor that we are going to drop. 
(This integral can be done by discretizing $\sigma_2$ and taking the step size
to zero at the end).  We must, however, keep the $\sqrt{G_{00}}$ contribution to
the $\sqrt{G}$ part of the path integral measure.  Since we are studying metrics of the
form\ (\ref{eq:staticmetric}), we can factor it out of the integral over $X^0$.

Replacing the result\ (\ref{eq:timeintegral}) in eq.(\ref{partfunct}) 
and performing the $\tau_1$ integration we obtain
\begin{eqnarray}
 Z & \simeq &  \frac{1}{\beta} \int^{\infty}  \,
   \frac{d\tau_2}{\tau_2}\,  
 \int_{X(0)=X(1)} \cD X(\sigma_2) \prod_{\sigma_2} \sqrt{G_{00} 
   \det G_{ij} (X(\sigma_2))} \times \\
 & \ & \ \ \ \ \times e^{-m_s^2 \int_0^1 d\sigma_2\,\left[ \frac{1}{2\tau_2}(G_{ij}\p_2 X^i \p_2 X^j) 
   + \tau_2 \left( \beta^2 G_{00}(X) - \beta_H^2 \right)\right]}\ .
\label{eq:rwpf}
\end{eqnarray}

By writing $\sigma_2 = \tau_2 t$, we can write this as:
\begin{equation}
  Z \simeq\ \frac{1}{\beta} \int_0^{\infty} \frac{d\tau_2}{\tau_2}
  \int_{X(\tau_2) = X(0)} \cD X \sqrt{G_{00}\det G_{ij}}
  e^{-\int_0^{\tau_2} dt L(X)}
\label{eq:newpf}
\end{equation}
where
\beq
L = \frac{m_s^2 }{2}G_{ij}\p_t X^i \p_t X^j + m_s^2 \beta^2 G_{00} - \beta_H^2\ .
\eeq
This is clearly an integral over random walks of the thermal scalar.
The paths are images under the modular transformation $\tau \to 1/\tau$
of configurations of strings at temperature $\beta$.

The path integral over $X_i$ can in principle be computed in the following
way.  The integral over $\cD X$ in\ (\ref{eq:newpf}) is the
heat kernel for the thermal scalar:
\begin{equation}
  K(\vec{X}_t, t; \vec{X}_0, 0) = \int_{X(0) = X_0,X(t)=X_t} 
  \cD X \sqrt{G_{00} G_{ij}} e^{- \int_0^t dt' L}
\label{eq:propagator}
\end{equation}
and solves the diffusion equation:
\begin{equation}
  - \frac{1}{\sqrt{G_{00} \det G_{ij}}} \p_i \sqrt{G_{00} \det G_{ij}}
  G^{ij} \p_j K + \left( \beta^2 G_{00} - \beta_H^2\right) K \equiv H K = -\p_t K
\end{equation}
with boundary conditions 
\begin{equation}
  \lim_{t\to 0} K(X_t, t; X_0,0) = \delta(X_t = X_0)
\end{equation}
Note the factors of $G_{00}$ in the wave equation; 
these arise from the explicit factors in the measure.
The thermal partition function for a single string can be written (assuming
it is dominated by large $\tau_2$) as:
\begin{equation}
  Z \simeq\ \frac{1}{\beta} \int_0^{\infty} \frac{d \tau_2}{\tau_2} \int dX_0
  K(X_0,\tau_2; X_0,0)
\label{eq:proptopf}
\end{equation}

We can write $K$ as a sum over eigenmodes of $H$:
\begin{equation}
  K = \sum_n A_n \psi_n(X_t,X_0) e^{-E_n t}
\label{eq:propexpand}
\end{equation}
this sum could be discrete or continuous.
If $H$ has a discrete spectrum with approximate spacing $\delta$ between
energy levels, then for $t \gg 1/\delta$, $K$ can be well-approximated
by the ground state wavefunction for the tachyon.  This is not the case for the string
in flat space -- $K$ then has a continuous spectrum, and the string
spreads without bound as $\tau_2 \to \infty$, as described in \S3.1.

\subsection{The radius of gyration for highly excited strings}

Given the propagator $K$ as defined in (\ref{eq:propagator}), we can
compute the typical size of a string as a function of the temperature $\beta$.
We take as this typical size
\begin{equation}
  (\delta x)^2 = \frac{1}{\beta Z} 
  	\int_0^{\infty} \frac{d\tau_2}{\tau_2} \int_0^{\tau_2} \frac{dt}{\tau_2} \langle \left(X(t) - X(0)\right)^2 \rangle(\tau_2)
\label{eq:rog}
\end{equation}
where
\begin{eqnarray}
	\langle \left(X(t) - X(0)\right)^2\rangle &=& \int d^d X_t \sqrt{G_{00}\det G_{ij}(X_t)}
		\int d^d X_0 \sqrt{G_{00} \det G_{ij}(X_0)} \nonumber \\
		& & \ \ \ \times K(X_0,\tau_2; X_t, t) (X_t - X_0)^2
		K(X_t,t; X_0,0)
\label{eq:rogtau}
\end{eqnarray}
As with the propagator itself, if the spectrum of $H$ is discrete with level spacing $\delta$,
and the integral in\ (\ref{eq:rog})\ is dominated by sufficiently large $\tau_2 \gg 1/\delta$, we can approximate
$K$ by the ground state wavefunction $A_0 \psi_0 e^{-E_0 t}$.  To see this, let us approximate $K$ by
the first two terms in\ (\ref{eq:propexpand}):
\begin{equation}
	K(X_2, t_2; X_1, t_1)  = A_0 \psi_0(X_2,X_1) e^{- E_0 (t_2 - t_1)}
		+ A_1 \psi_1(X_2,X_1) e^{-E_1 (t_2 - t_1)} + \ldots
\end{equation}
We can then perform the integral over $t$ in\ (\ref{eq:rogtau}):
\begin{eqnarray}
	& & \int_0^{\tau_2} \frac{dt}{\tau_2} K(X_0,\tau_2; X_t, t) K(X_t,t; X_0,0) =
	A_0^2 \psi_0(X_0,X_t)\psi_0(X_t,X_0) e^{-E_0\tau_2} \nonumber \\
	& & \ \ \ \ \ \ \ + A_1^2
		\psi_1 (X_0,X_t)\psi_1(X_t,X_0) e^{-E_1 \tau_2} \nonumber \\
		& & \ \ \ \ \ + \frac{A_0 A_1}{\tau_2(E_1 - E_0)} \left(
			e^{-E_0 \tau_2} - e^{-E_1\tau_2} \right)
		 \nonumber \\
		& & \ \ \ \ \ \ \ \ \ \ \ \ \ \ \ \ \times \left( \psi_0(X_0,X_t)\psi_1(X_t,X_0) + \psi_1(X_0,X_t)\psi_0(X_t,X_0)\right)
\end{eqnarray}
The first term clearly dominates when $\tau_2 \gg \frac{1}{E_1 - E_0}$.  In this case, the radius of
gyration $r_g = \sqrt{(\delta x)^2}$ can be well approximated by the width of the ground state
wavefunction $\psi_0$.

\subsection{Excited strings in Anti-de Sitter space}

We would like to study the results of the previous section for the spacetime
$AdS_D \times S^n$, where each factor has radius of curvature $R$,
as we would find in Freund-Rubin-type compactifications of supergravity
\cite{Freund:1980xh}.  We will assume $R \gg \ell_s$, so that
the worldsheet sigma model is weakly coupled.
The metric for this spacetime is
\begin{equation}
	ds^2 = - \left(1 + \frac{r^2}{R^2}\right) dt^2 + \frac{dr^2}{\left(1 + \frac{r^2}{R^2}\right)}
	+ r^2 d\Omega_{D-2}^2 + R^2 d\Omega_n^2
\end{equation}
where $d\Omega_k^2$ is the metric for a $k$-sphere with unit radius.  Note that
these backgrounds in string theory typically have Ramond-Ramond tensor
fields.  While this implies a far more complicated worldsheet action than the one
we are studying, we will ignore this effect and focus on the worldsheet action for the bosonic modes.

In this case, the propagator $K$ satisfies the wave equation
\begin{eqnarray}
	&& \frac{-1}{r^{D-2}}\p_r \left(r^{D-2}  \left(1+\frac{r^2}{R^2}\right)
	\right)\p_r K  -  \frac{\del^2_{D-2}}{r^2}K  \nonumber \\
	& &  \ \ \ \ \ \ \ \ \ \ \ \ - \frac{\del^2_n}{R^2} K+
	\frac{\beta^2}{\ell_s^4} \left(1 + \frac{r^2}{R^2}\right)K  - \frac{\beta^2_H}{\ell_s^4} K
	= - \p_t K
\label{eq:frwaveeqn}
\end{eqnarray}
where $\del_k^2$ is the Laplacian on the unit $k$-sphere.  This equation is separable:
if $\theta^a$ denote the angular coordinates in $AdS_D$ and $\psi^m$ the
angular coordinates along $S^n$, then the solution to\ (\ref{eq:frwaveeqn})
satisfying the required boundary conditions is:
\begin{equation}
	K(r,\theta,\psi) = K_{D}(r,\theta,t) K_{n}(\psi, t)
\end{equation}
where
\begin{eqnarray}
	&& H_D K_D = \frac{-1}{r^{D-2}}\p_r \left(r^{D-2} \left(1+\frac{r^2}{R^2}\right)
	\right)\p_r K_D  -  \frac{\del^2_{D-2}}{r^2}K_D \nonumber \\
	&&\ \ \ \ \ \ \ \ \ \ + \frac{\beta^2}{\ell_s^4} \left( 1+ \frac{r^2}{R^2}\right) K_D
	- \frac{\beta_H^2}{\ell_s^4} K_D = - \p_t K_D
\label{eq:adsproppart}
\end{eqnarray}
and
\begin{equation}
	H_n K_n - \frac{\del_n^2}{R^2} K_n = - \p_t K_n
\label{eq:sphereheat}
\end{equation}

Even without the additional tachyon mass term, the spectrum of
the Laplacian in the above coordinates (covering all of AdS)
is discrete.  The tachyon wave equation, with the additional mass term
will certainly have a discrete spectrum as well.  Therefore, from the standpoint of string
thermodynamics, this spacetime behaves as if the spatial directions are completely
compactified -- the power-law prefactor $\alpha$ in the density of states 
$$ \rho(E) = \frac{e^{\beta_H E}}{E^{1+\alpha}} $$
arises from target space directions which leads to a continuous spectrum of
eigenvalues of the tachyon wave operator.  In the case of $AdS_D$ spacetimes, we
expect that $\alpha = 0$ for energies high enough that the curvature of
$AdS_D \times S^n$ begins to affect the string spectrum.

We can think of anti-de Sitter space as a hyperbolic space with a gravitational
potential.  Random walks in hyperbolic space tend to move towards the boundary
at infinity \cite{Monthus:1996}, which is entropically favored due to the large volume there.
However, we will see that the gravitational potential overwhelms this tendency,
so that the radius of gyration for random walks asymptotes to a finite value.

For sufficiently small temperatures and energies, we expect the string to act like
a string in flat space -- its size $r_g = \sqrt{(\delta x)^2} \sim \sqrt{(\ell_s^3 M)}$
scales with the square root of the length $L \sim \ell_s^2 M$, 
where $M$ is the mass of the string state.  As the temperature and thus
the average energy $M$ increases, the string will spread.  The integral over the modular
parameter $\tau_2$ has a saddle point at roughly $1/M$ (more precisely,
under a modular transform $\tau = -1/\tau'$, the paths of the thermal scalar become
string configurations, and $\tau'  M$ where $M$ is the mass of the
excited string state).  We will find that there are two characteristic temperatures
at which the string starts to feel the curvature.  At the lower temperature
$T_1$, the dominant values of $\tau_2$ in\ (\ref{eq:rog})\ are long enough that
$K_D$ is well-approximated by its ground state wavefunction, and the
string stops spreading in the $AdS$ direction.  At a higher temperature $T_2$ the string
spreads over the entire $S^n$, and $r_g$ can increase no further.

$K_n$ satisfies the heat equation on $S^n$: solutions
can be found, for example, by analytic continuation of
the results in \cite{Dowker:1975xj,Dowker:1975tf} to Euclidean space.
If $s$ is the geodesic distance on $S^n$ between $X_t$ and $X_0$, and $R$ the radius,
and $p = \cosh \frac{s}{R}$, 
\begin{equation}
	K(X_t,X_0; t) = \frac{1}{\left(2\pi R^2\right)^{\frac{n}{2}}}
	\left( \frac{d}{dp}\right)^{\frac{n}{2}-1} \sum_{k=0}^{\infty}
		P_k(p) \exp\left\{- \frac{t}{R^2} \left((k+\half)^2 - (n - \half)^2\right)\right\}
\end{equation}
where $P_n$ are the Legendre polynomials.
Note that the eigenvalues of $\p_t$ have spacing of order 
$$\delta_n \sim 1/R^2\ , $$
as implied by\ (\ref{eq:sphereheat}).

$K_D$ cannot be solved exactly.  However, we can estimate the level spacing for
the lowest eigenvalues of $\p_t$, and the width of the corresponding wavefunctions,
using scaling arguments.  Begin by defining dimensionless
variables $x = \frac{r}{R}$.  The eigenvalue equation for the left hand side of\ (\ref{eq:adsproppart})\
is:
\begin{equation}
	\frac{1}{R^2 x^{D-2}}\p_r \left(x^{D-2}(1 + x^2)\right)\p_r K - \frac{\del_{D-2}^2}{R^2 x^2} K
	+ \frac{\beta^2}{\ell_s^4} x^2K  + \frac{\delta\beta^2}{\ell_s^4} K = E K
\end{equation}
where $\delta \beta^2 = \beta^2 - \beta_H^2$.  Let us multiply the whole equation by $R^2$.
Defining dimensionless parameters $\cE = R^2 E$,
\begin{equation}\label{eq:waveoperatorev}
	\mu = \cE - \frac{\delta\beta^2 R^2}{\ell_s^4}\ ,
\end{equation}
and
$$ \alpha^2 = \frac{\beta^2 R^2}{\ell_s^4} $$
we find that the eigenvalue equation is completely controlled by $\alpha$:
\begin{equation}
	- \frac{1}{x^{D-2}} \p_x \left(x^{D-2}(1+x^2)\right)\p_x K 
	- \frac{\del^2_{D-2}}{x^2} K + \alpha^2 x^2 K = \mu K
\label{eq:dimlessev}
\end{equation}

In the cases we are interested in, we expect $\alpha$ to be large.  That is, we are interested
in string theory on spaces for which $R/\ell_s \gg 1$.  We may also tune $\beta$.  However,
we are interested in temperatures high enough so that the radius of gyration is
significantly affected by the AdS background.  We will find that such temperatures are
(self-consistently) very close to $\beta_H^{-1} \sim m_s$.

In the limit of large $\alpha$, the quadratic "potential" term in\ (\ref{eq:dimlessev})\ 
should ensure that the eigenfunctions have a small width.  If we define
$x = \frac{y}{\sqrt{\alpha}}$, we can write the first term in\ (\ref{eq:dimlessev})\ as
\begin{equation}
	\frac{\alpha}{y^{D-2}} \p_y \left(y^{D-2} \left(1 + \frac{y^2}{\alpha}\right)\right)\p_y K
	\simeq \frac{\alpha}{y^{D-2}} \p_y y^{D-2}\p_y K
\end{equation}
Therefore, we can rewrite\ (\ref{eq:dimlessev}) as
\begin{equation}
	\cH K = - \frac{1}{y^{D-2}} \p_y y^{D-2}\p_y K - \frac{\del_{D-2}^2}{y^2}K
	+ y^2 K = \frac{\mu}{\alpha} K
\end{equation}
In these units, we can expect that the low-lying eigenfunctions of $\cH$ have width
$(\delta y)^2 \sim 1$ and eigenvalues $\mu/\alpha \sim 1$.  Translated into
our original dimensionful variables, we find that the width of the low-lying
eigenfunctions of $H_D$ is approximately:
\begin{equation}\label{eq:adsradgyr}
	(\delta r)^2 \sim \frac{R^2}{\alpha} = \frac{\ell_s^2 R}{\beta}
\end{equation} 
while the energy eigenvalues are of the order
\begin{equation}\label{eq:escaling}
	E \sim \frac{\alpha}{R^2} + \frac{\delta\beta^2}{\ell_s^4} = \frac{\beta}{\ell_s^2 R} +
		\frac{\delta\beta^2 }{\ell_s^4}
\end{equation} 
and the gap between energy levels can be expected to be of order
$$ \delta_D \sim \frac{\alpha}{R^2} = \frac{\beta}{\ell_s^2 R} \ . $$
We can see from (\ref{eq:adsradgyr})\ a major difference between long strings in
AdS space and flat space.  In the latter case, the radius of gyration would diverge
as $\beta \to \beta_H$ \cite{Horowitz:1997jc}.

We expect that the picture of single strings at finite temperature is as follows.
At low enough temperatures, the strings are small and do not feel the curvature
of $AdS_D\times S^n$. The radius of gyration grows as the square root of their mass:
$r_g^2 \sim \ell_s^3 M$.  For highly massive strings in flat space, the integral over
the modular parameter $\tau_2$ has a saddle point at $\tau_2 \sim \frac{1}{\ell_s M}$.
After the modular transform to the thermal scalar picture, and rescaling, this
becomes $\tau_2 \sim M$.  In the canonical ensemble for a single string in flat space,
we can write (see for example \cite{Horowitz:1997jc}):
\begin{equation}
	M = \frac{\ell_s}{\delta \beta^2} 
\label{eq:mtobeta}
\end{equation}
Therefore the string becomes massive and large as $\beta$ approaches $\beta_H \sim \ell_s$.
When the temperature grows to temperature $T_1$ such that
$M \sim \frac{R}{\ell_s \beta}$, the integral over $\tau_2$ can be expected to have
a saddle point at $\tau_2 \sim \ell_s^3 M = r_g^2 \sim \frac{\ell_s^2 R}{\beta}$.   
At this point, $\tau_2$ is of the order of the level spacing $\delta_D$ for
eigenvalues of the operator $H_D$.  For higher temperatures,
therefore, we expect that $K$ is well-approximated
by the ground state wavefunction and the string ceases to grow in the $AdS_D$ direction.
We should still be able to use\ (\ref{eq:mtobeta}) to estimate the temperature,
as the flat space approximation is only just starting to fail: we find that
$\delta \beta^2 \sim \frac{\ell_s^3}{R} \ll \ell_s^2$; the temperature is quite close to the Hagedorn
temperature.

The string has stopped expanding in the radial direction of $AdS_D$ because
of the gravitational potential.  However,
the string can still expand in the $S^n$ direction.  We expect the expansion of
the string in this direction to be essentially that of a string in $\R^n$, until the
size of the string is of order the $S^n$ radius $R$.  We expect that at this point,
the integral over $\tau_2$ is dominated by values of order $M \ell_s^3 \sim r_g^2 \sim R^2$,
and $K_n$ is well-approximated by the constant mode on $S^n$.

At this point the string feels the curvature in all spatial directions, and
for all intents and purposes sees the spacetime as being completely compactified.
As discussed at the end of \S2, single string dominance will cease at this point, and
multiple-string states will start to become equally likely.

One might have wondered whether the thermal scalar would lead
to an instability above some temperature $T_H = \beta_H^{-1}$.
AdS space allows for tachyons that do not indicate instabilities 
\cite{Breitenlohner:1982bm,Breitenlohner:1982jf}.   Furthermore,
in AdS space at finite temperature, the proper radius of the
Euclidean time direction grows as one approaches the boundary of AdS.
Thus the mass of the thermal scalar gets large and positive at large radius, after the fashion of the
"localized" and "quasi-localized" tachyons discussed in \cite{Adams:2001sv,
Adams:2005rb, Horowitz:2005vp}, so that the tachyon is at best localized in the center of
AdS space \cite{Barbon:2001di}.  However, so long as
a ground state wavefunction exists with the scaling properties we have described,
we are guaranteed such an instability.  For fixed $\mu,\alpha$, (\ref{eq:dimlessev})\ 
is independent of the temperature, and a lowest value of $\mu/\alpha$ will exist
and be of order $1$.  However, the relationship between $\mu$ and $E$ is temperature-dependent,
leading to the statement (\ref{eq:escaling}).  It is clear that $E$ will become negative
when $\beta^2 \sim \beta_H^2 - \O(\ell_s^2/R)$: the Hagedorn temperature will
be raised by a factor of order $1/R$.  

It might be tempting to speculate that for $R \ll \ell_s$, this tachyon is lifted.  
In the case of $AdS_5\times S^5$ compactifications, this regime is dual to
a weakly coupled gauge theory.  A similar story was
proposed for the tachyon in type 0 theories \cite{Klebanov:1998yy,Klebanov:1999ch,
Klebanov:1999um}. On the other hand, in those theories, there is evidence
that the instability persists in the weakly-coupled large-N gauge theory
\cite{Tseytlin:1999ii,Adams:2001jb}\ as a Coleman-Weinberg-type instability.
In the case of large-N field theories at finite temperature on $S^3$
(dual to $AdS_5$ in global coordinates), the Hagedorn
transition is also known to exist at weak coupling \cite{Aharony:2003sx,
Aharony:2005bq, Alvarez-Gaume:2005fv}.

\section{The Hagedorn transition in anti-de Sitter spacetime}

In this section we will study the Hagedorn transition in AdS spacetime.
In flat space, above the Hagedorn temperature, the
thermal scalar
becomes a tachyon.  One expects the high-temperature phase
on Euclidean space to correspond to a vev for this scalar.  This is analogous
to the statement \cite{Polyakov:1978vu,Susskind:1979up}
that the Polyakov-Susskind loop $P\tr e^{-\oint A_0 dx^0}$
(or its norm squared \cite{Aharony:2003sx})
gets an expectation value in the deconfined phase of gauge theories.  
In string theory, this transition is highly problematic in flat space -- as discussed
in the introduction and in \cite{Atick:1988si}, the canonical ensemble
does not strictly exist for interating strings even below the Hagedorn temperature;
while above the Hagedorn temperature, the condensation involves free energies
of order the inverse gravitational coupling, and computational control is lost.

However, for string theory in anti-de Sitter spacetimes the AdS curvature provides
a "box"  in which to study string theory, and provides a dual field theory description for
which one expects thermodynamics to be well defined.
For example, as we will argue below following refs \cite{Witten:1998zw, Barbon:2001di, 
Barbon:2004dd, Adams:2005rb}, 
the analogy between the thermal scalar and the Polyakov-Susskind loop
can be made precise in anti-de Sitter spacetimes which
are dual to (conformal) gauge theories. (These conformal field theories do not confine,
but the large 't Hooft coupling theories do exhibit charge screening at high temperatures
\cite{Maldacena:1998im}, and the theories at weak and strong coupling do see a jump in their
free energies from ${\cal O}(1)$ to ${\cal O}(N^2)$ \cite{Witten:1998qj,Witten:1998zw,
Aharony:2003sx,Aharony:2005bq}.)  This is related
to the statement, discussed in \cite{Barbon:2001di,Barbon:2002nw,Barbon:2004dd} 
that the natural endpoint of the condensation process is the Euclidean black hole.

In anti-de Sitter space
at the Hagedorn transition, the wavefunction for the tachyonic thermal 
scalar is localized near the origin, where the mass is smallest.  The condensation
of localized tachyonic winding modes has been studied recently \cite{Adams:2005rb}\
with the conclusion that the winding direction pinches off.  As we will discuss,
in thermal AdS space times this is consistent with the endpoint being the AdS black hole.

\subsection{Endpoints of the condensation of the thermal scalar}

The following picture of the thermodynamics of string theory on global $AdS_5$
appears to hold at weak and strong coupling \cite{Witten:1998qj,Witten:1998zw,
Aharony:2003sx,Aharony:2005bq,Alvarez-Gaume:2005fv}.  As the temperature
reaches $1/R$, large and small black holes become extrema of the
free energy.  At a higher temperature also of order $1/R$, the 
theory undergoes a first-order Hawking-Page phase transition
\cite{Hawking:1982dh} and a "large" black hole with horizon radius $r_s > R$
becomes thermodynamically preferred. The "small" black hole solution 
has horizon radius less than $R$; as one increases the temperature beyond
the Hawking-Page transition temperature,
its free energy is larger than that of
thermal $AdS$ or of the large AdS-Schwarzchild black hole.  
As argued by \cite{Barbon:2001di,Barbon:2002nw,
Barbon:2004dd,Alvarez-Gaume:2005fv}, at temperatures of order the Hagedorn
temperature, the small black hole and thermal AdS solutions coalesce.  At these
temperatures the horizon of the small black hole will have curvatures of order the
string scale, and should undergo a transition to a long fundamental string
\cite{Horowitz:1996nw}.

\begin{figure}
\centering
\epsfig{file=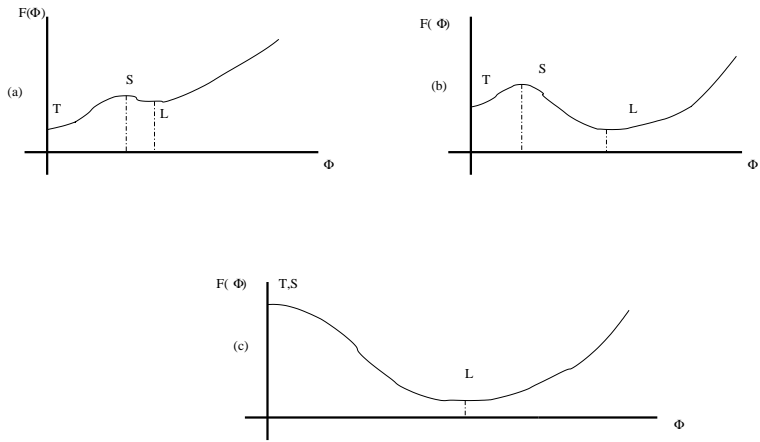, width=10cm}
\caption{{\it The free energy $F(\Phi)$ as a function of the log of
the norm of $W(C)$,
as the temperature increases.  The labels T,S, and L label the extrema of the free energy
corresponding to "thermal AdS", the "small" black hole, 
and the "large" black hole, respectively.
Figure (a) represents the free energy at a temperature
just above that at which the black hole solutions begin to exist.  Figure (b) represents
a temperature above the Hawking-Page transition temperature, for which the large
black hole is thermodynamically stable. Figure (c) represents the Hagedorn transition,
at which $T$ and $S$ merge and thermal AdS becomes a local maximum of the free energy.}}
\label{fig:freecurve} 
\end{figure}

The natural picture of the free energy is shown in Fig. \ref{fig:freecurve}.  The order parameter
shown is related to the Polyakov-Susskind loop in the
gauge theory.  More precisely, we want to compute something like
the vev of the norm of the "Maldacena-Wilson" loop that
includes the adjoint scalars in the $\CN=4$ vector multiplet:
\begin{equation}\label{eq:mwilson}
	W(C) = \tr \exp\left( i \oint dt \left( \dot{x}^\mu(t) A_\mu - i y_i(t) \Phi^i \right)\right)\ .
\end{equation}
This can be computed by studying worldsheets wrapping the
thermal circle in Euclidean $AdS_5\times S^5$ \cite{Maldacena:1998im,Rey:1998ik,Rey:1998bq,
Drukker:1999zq}.  
It is important that we compute the norm
\cite{Aharony:2003sx}: in finite volume, one should integrate over the phase of the
$W(C)$, leading to a vanishing expectation value 
\cite{Witten:1998zw,Aharony:2003sx}.  
The statement in the bulk dual at strong coupling is that one should integrate over
the value of the NS-NS B-field through the "cigar" parameterized by $t,\rho$
\cite{Witten:1998zw}.

We would like to relate a nonvanishing vev for the norm of $W(C)$ to
a condensate of thermal scalars in the bulk.  The basic argument is as follows.
The expectation value of $W(C)$ in the gauge theory is dual to
the classical action of a fundamental string worldsheet which wraps the Euclidean
time direction in the bulk \cite{Witten:1998zw,Maldacena:1998im,Rey:1998ik,Rey:1998bq,
Brandhuber:1998bs,Brandhuber:1998er}.\footnote{The calculations listed in the references
above were done in Poincar\'e coordinates, and aside from \cite{Witten:1998zw} were
concerned with Wilson loops or with computing the quark-atiquark potential.
A calculation of this potential in global AdS can be found in \cite{Ebrahim:2005uz}.}
In AdS spacetimes, no such classical solution exists.  In the presence of a condensate of winding
modes, however, the worldsheet can end at the condensate, since a condensate of
winding modes spontaneously breaks winding number.

On the other hand, at finite temperature the
the expectation value of this loop can be calculated in the dual supergravity
backgrounds corresponding to extrema of the free energy, without reference
to the thermal scalar.  The existence of nontrivial solutions arises
in the presence of black holes, which change the topology
of the spacetime so that a circle along the Euclidean time direction can be
contracted to a point at the black hole horizon.

We can show that at strong coupling, $\ln \sqrt{\vev{|W|^2}} = r_s$ for a black hole
with radius $r_s$. One performs the bulk computation of $W(C)$by computing the
action of a string worldsheet which wraps the $t,\rho$ plane at fixed angular
coordinate 
\cite{Witten:1998zw,Maldacena:1998im,Rey:1998ik,Rey:1998bq,Brandhuber:1998bs,Brandhuber:1998er}\ 
and which asymptotes to the curve
$C$ at the boundary of $AdS$.
\footnote{In principle our calculation is too quick.
For example, we are ignoring $y(t)$ in (\ref{eq:mwilson}).
In general, expectation values and correlators of $W(C)$ 
have a linear divergence unless $\dot{x}^2 = \dot{y}^2$
\cite{Drukker:1999zq}.
Nonetheless, we are interested
in the behavior of the string worldsheet deep in the interior
of the AdS and AdS-Schwarzchild geometries.  Therefore we expect our calculation,
with our cutoff, to be qualitatively correct.}

To compute the norm squared of $W(C)$,
we will compute the correlator of two such loops placed on the antipodes of the
spatial $S^3$ in the gauge theory.  At large $N$, the disconnected part of the
correlator should dominate, so that 
\begin{equation}
    \vev{W(C) W^{\dagger}(\tilde{C})} \equiv \vev {|W(C)|^2}
\end{equation}
where if $C$ is a curve along Eulcidean time and localized at a point $x\in S^3$,
then $\tilde{C}$ is the curve along Eucildean time localized at the antipode
of $x$ in the $S^3$.

In general these worldsheets are infinite due to the large-$\rho$
limit, corresponding to UV divergences in the field theory calculation
\cite{Maldacena:1998im,Drukker:1999zq}.  We will regulate these by subtracting
the contribution to $\vev{W(C_1)W(C_2)}$ in thermal AdS.\footnote{Again,
see the caveats in the previous footnote}.
The result is that the regulated correlator vanishes for the thermal
$AdS$ contribution, reflecting the fact that the topology of thermal
AdS forbids a classical worldsheet contributing to $\vev{W(C)}$ even
before integrating over the NS-NS B field.

The metric of both AdS space and the AdS black hole can be written as:
\begin{equation}
 ds^2 =  f(r) dt^2 + \frac{1}{f(r)} dr^2 + r^2 d\Omega_3^2
\end{equation}
For AdS space, $f(r) = 1 + \frac{r^2}{R^2}$.  For the $AdS_5$ black hole with mass $M$,
\begin{equation}
    f(r) = 1 + \frac{r^2}{R^2} - \frac{8 G_N M}{3\pi r^2} = 
    1 + \frac{r^2}{R^2} - \frac{w M}{r^2}
\end{equation}
Here $G_N \sim \ell_{p,5}^3$ is the five-dimensional Newton's constant.
$\ell_{p,5}$ is the five-dimensional Planck length.  Note that
$R^5 \ell_{p,5}^3 = \ell_{p,10}^8$, where $\ell_{p,10}$ is the
ten-dimensional Plank scale.

The Schwarzchild radius $r_h$, the location of the horizon in these coordinates, 
is the largest root of $f = 0$,
\begin{equation}
    r_h^2 = \frac{R^2}{2}\left( - 1 + \sqrt{1 + \frac{4wM}{R^2}}\right)
\end{equation}
Note that for $\ell_{p,5}^3 M \ll R^2$, $r_h^2 \sim 2\ell_{p,5}^2 M \ll R^2$.  The
black hole is much smaller than the AdS radius of curvature and is well-approximated
by the 5d Schwarzchild black hole in flat space.  For $\ell_{p,5}^3 M \gg R^2$, 
$r_h^2 \sim 2 R \sqrt{wM}$ and the black hole is a 5d AdS-Schwarzchild black hole
with size much larger than the AdS radius.

The corresponding temperature is set by the requirement that the metric be nonsingular when
$f = 0$.  This leads to:
\begin{equation}
    \beta = \frac{4\pi R^2 r_h}{4r_h^2 + 3 R^2}
\end{equation}
Solving for $r_h$ we find two radii for a given temperature:
\begin{equation}
    r_h = \frac{\pi R^2}{2\beta} \left( 1 \pm \sqrt{1 - \frac{3\beta^2}{\pi^2 R^2}}\right)
\end{equation}
which correspond to the "small" and "large" black holes with radii smaller
and larger, respectively, than the AdS radius. Note that $r_h$ represents
the minimum of the coordinate $r$.  When $\beta \ll R$,
well above the Hawking-Page transition temperature,
\begin{equation}
    r_h = \frac{\pi R^2}{2\beta}\ \ \ \ \ {\rm or}\ \ \ \ \ \frac{3\beta}{2\pi}
\end{equation}
for large and small black holes, respectively.  

We compute the action of the string wrapping the $r,t$ coordinates by computing the
Nambu-Goto action.  Let the worldsheet coordinates be $\tau,\sigma$ so that
$t = \tau$, $r = \sigma$.  Then
\begin{equation}
    S = \int_0^{\beta} d\tau \int^{r_{max}}_{r_h} d\sigma = \beta(r_{max} - r_h)
\end{equation}
where $r_{max}\to\infty$ is a cutoff on the worldsheet action, representing a UV cutoff
in the definition of the $W(C)$ \cite{Maldacena:1998im}.  Similarly, the
contribution of thermal AdS to the correlator of two such loops placed on 
antipodes of the $S^3$ is just $2\beta r_{max}$. We can subtract half of this to find that
$\Phi \equiv \ln \sqrt{\vev{|W|^2}} = 0, r_{h,small}(\beta), r_{h,large}(\beta)$ for thermal AdS, 
the small black hole, and
the large black hole, respectively.  In the high temperature limit $\beta \gg R$, 
\begin{eqnarray}
    \Phi & =& 0 \ \ \ \ \ {\hbox{for thermal AdS}}\cr
    &=& \frac{3}{2\pi T} \ \ \ \ \ {\hbox{for the "small" black hole at temperature T}}\cr
    &=& \pi R^2 T \ \ \ \ \ {\hbox{for the "large" black hole at temperature T}}
\end{eqnarray}
justifying the identification by \cite{Aharony:2003sx,Aharony:2005bq,Alvarez-Gaume:2005fv}
of the maxima and minima of the free energy diagram seen in Fig. \ref{fig:freecurve}.

As we discussed above, we also expect a non-zero expectation 
value for $\Phi$ if the thermal scalar condenses.  A worldsheet stretching from $C$ on the boundary
along $r,t$ would extend along the $r$ direction.  At a point where it met the condensate,
the worldsheet would be able to end, due to the presence of winding modes.

To see how the supergravity calculation of $\Phi$ and the
thermal scalar picture might be related, consider at the Hagedorn temperature
for string theory in AdS space.  As noted by \cite{Alvarez-Gaume:2005fv}, at temperatures
of order the string scale, the "small black hole" solution has a horizon of order the string
scale.  This is the Horowitz-Polchinski correspondence point \cite{Horowitz:1996nw},
and at higher temperatures the black hole is expected to be better described
as a long string.  On the other hand, the "thermal AdS" solution corresponds to
a gas of strings and supergravity particles at finite temperature: close to the
Hagedorn transition, the dominant configurations are single very long strings
\cite{Mitchell:1987hr,Mitchell:1987th,Bowick:1989us}.  It is reasonable to
conjecture that at this temperature the local maximum of the free energy $F(\Phi)$
corresponding to the "small black hole" merges with the metastable minimum corresponding
to thermal AdS space.  This would correspond to a tachyonic instability in $F(\Phi)$.
At the same time, one expects the thermal scalar to become a quasilocalized tachyon
in AdS space.  A natural guess for the endpoint of this condensation process is the
big black hole \cite{Barbon:2001di}.  

\begin{figure}
\centering
\epsfig{file=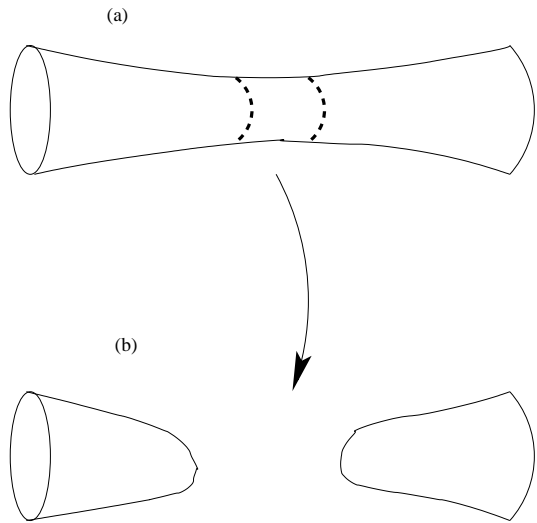, width=7cm}
\caption{{\it A cylinder with varying radius.  The region between the heavy black lines
in (a) represents a region for which strings winding around this circle become
tachyonic.  Figure (b) represents the conjectures endpoint of the condensation
of this tachyon.}}
\label{fig:pinch} 
\end{figure}

Adams {\it et. al}\ \cite{Adams:2005rb}\ have studied quasilocalized tachyons corresponding to
winding modes on cylinders with antiperiodic boundary conditions
for fermions, whose radius becomes smaller than $\ell_s$ in some region of the cylinder.
They have argued that the condensation of the tachyon leads to the "pinching off"
of the cylinder,m as shown in Fig. \ref{fig:pinch}.  This is consistent with the well-known
RG flow of the 2d Gaussian (or "XY") model on a circle above the Kosterlitz-Thouless
transition \cite{Kosterlitz:1973xp,Kosterlitz:1974sm}.  The winding modes
of the string around the circle are the vortices of the XY model, and as the vortices
condense, the radius of the circle is driven to zero.  Furthermore, according to
\cite{Adams:2005rb,McGreevy:2005ci}\ the tachyon
potential on the worldsheet suppresses fluctuations of the string in the
region where the tachyon has condensed, much like the worldsheet description
of D-brane decay in \cite{Harvey:2000na}.

We therefore expect that below the Hagedorn transition, 
the free energy curve $F(\Phi)$ shown in Fig. \ref{fig:freecurve}\ can be swept out by "turning on"
the thermal scalar.  In thermal AdS it is an  irrelevant operator on the worldsheeet, 
but in the presence of the "small" black hole, the tachyon 
becomes a relevant operator, and can induce an
RG flow either back to thermal AdS or to the "big" black hole.
Heuristically, a worldsheet wrapping the thermal circle
and extending in the bulk can end on the condensate of winding modes.
On the other hand, we know that in black hole backgrounds, the vev
$\langle W(C)W(\tilde{C}) \rangle$ is nonzero because the thermal circle pinches off at the
horizon.  By the discussion in the previous paragraph, we expect that these two
mechanisms are somehow dual to each other, and in Schwarzchild
coordinates the physics of the black hole horizon is related to the physics of the
thermal scalar.\footnote{Horowitz \cite{Horowitz:2005vp}\ suggests that for extremal
black holes, the thermal scalar condenses near the horizon and modifies the geometry,
providing a possible reconciliation between the string theory
\cite{Strominger:1996sh}\ and supergravity \cite{Hawking:1994ii}\ calculations
of the entropy of an extremal black hole.}

Previous work has also indicated that the horizon in supergravity is related to the tachyon
condensate.  As Dabholkar \cite{Dabholkar:2001if} has discussed, 
following earlier insights by Susskind and Uglum \cite{Susskind:1994sm},
this may help us understand the Bekenstein-Hawking entropy of the black hole.
Dabholkar begins with the following  observation from \cite{Susskind:1994sm}.
The near-horizon geometry of the Schwarzchild black hole is Rindler space,
\begin{equation}
    ds^2 = -\rho^2 dt^2 + d\rho^2 + \ldots
\end{equation}
Upon Euclidean continuation, we find that $t\sim t+2\pi$ is required
to avoid a conical deficit angle.  This means that Euclidean time
has a periodicity equal to the temperature of the black hole.  Now the
entropy should be $S = \beta^2\partial_{\beta} F$, where $F$ is the free energy
computed in this background.  To vary theta, one must vary the periodicity of
$t$ and so introduce a conical deficit angle.  Dabholkar computed the free
energy as a function of this angle by computing the free energy for the
nonsupersymmetric orbifold $\mathbb{C}/\mathbb{Z}_N$, which has
a conical definit angle $2\pi\left(1-\frac{1}{N}\right)$  These orbifolds
have twisted sector tachyons, inducing decays from 
$\mathbb{C}/\mathbb{Z}_N \to \mathbb{C}/\mathbb{Z}_{N-2}$.
The free energy of these orbifolds
scale linearly with $N$.  By analytically continuing in $N$
one computes $S = \beta^2\partial_{\beta} F$ exactly reproducing the tree-level 
Bekenstein-Hawking entropy.

The tachyons inducing decay of the orbifold \cite{Adams:2001sv}
are concentrated at the tip of the cone.  In the light of the above description of
the "pinching off" of AdS space to form a Euclidean black hole, via condensation
of the thermal winding modes, it is natural to identify these localized tachyons as being excitations
of the winding mode condensate.  The winding modes induced the pinching off
of the thermal circle, so it is natural that shifting this condensate should
change the geometry of the cones along the lines of \cite{Adams:2001sv, Dabholkar:2001if}.

In Lorentzian signature, the thermal scalar has a less direct interpretation.
The results of \S2, \S3 indicate that the dominance of the
thermal scalar in the partition function near the Hagedorn
transition is dual in some sense to the dominance of long strings
in the microcanonical ensemble.  We speculate that the appearance of
closed string tachyons in computing the black hole entropy is 
a sign that the entropy is accounted for by long strings near the horizon,
after the fashion of \cite{Susskind:1994sm}.

\subsection{A conjecture regarding the tachyon potential}

In the previous section, following refs. \cite{Barbon:2001di, Barbon:2002nw,
Aharony:2003sx,Aharony:2005bq,Alvarez-Gaume:2005fv}, we pointed out 
that the free energy diagram for
finite-temperature strings in anti-de Sitter space could be written
as a function of the expectation value of a variant of the Polyakov-Susskind loop, and that this
could be related to the bulk expectation value of the thermal scalar.  In particular,
at the Hagedorn transition, the free energies of the small black hole and of the
long string dominating the entropy of thermal AdS merge.  The free energy 
of thermal AdS as a function of the norm of the vev of $W(C)$ becomes
a local maximum precisely at the point the thermal scalar in the bulk becomes
tachyonic.  The natural endpoint of the condensation of the thermal scalar is the
large black hole, and there is some evidence that the tachyon accounts for the
Bekenstein-Hawking entropy.

W will use this conjecture to  estimate the potential energy of the tachyon condensate
(assuming that such a potential makes sense), by demanding that it
reproduce the free energy of the large black hole.
The scaling of this dependence with $R_{ads},\ell_{p,10}$, where $R_{ads}$ is the
radius of curvature of $AdS_5\times S^5$, will imply
a coupling between the tachyon potential and the spacetime Ricci scalar.

The free energy difference between thermal AdS and the "big" black hole is:
\begin{equation}
  I = \beta F \propto - \frac{r_h^3}{\ell_{p,5}^3}
\end{equation}
in the limit $r_h \gg R_{ads}$.  We propose to equate this to the free energy difference
due to the thermal scalar potential energy $V(\phi)$ .  We will assume that inside
a region of the size $R^5 r_h^4$ in nine spatial dimensions (we are including
the $S^5$ factor), the tachyon potential $V$ is at its minimum,
and outside this region it is at its maximum.  The difference $\delta V$ then scales
as
\begin{equation}\label{eq:tachyonpotinads}
  \delta V \sim \frac{1}{R_{ads}^7 \ell_{p,5}^3} \sim \frac{1}{R_{ads}^2 \ell_{p,10}^8}\ .
\end{equation}
The dependence on $\ell_{p,10}^8 = g_s^2 \ell_s^8$, with $g_s$ the ten-dimensional
string coupling and $\ell_s$ the string scale, is correct if this potential
is generated at tree level in closed string theory.  The scaling with $g_s^2$ of this 
tree-level free energy is consistent with the results in \cite{Atick:1988si}.  The dependence
on the AdS radius is puzzling.  In particular $\delta V$ seems to
disappear in the flat space limit.  In itself this latter point should be
related to the fact that the ``large'' AdS black hole isn't a solution in
flat space.  On the other hand, one might have expected a term which scales as
$1/{g_s^2 \ell_s^{10}}$ corresponding to the tachyon potential in flat space.

The scaling (\ref{eq:tachyonpotinads}) will be arise if there is a coupling
\begin{equation}\label{eq:tachact}
  S_{tachyon} = \int d^{10}x \sqrt{g} V(\phi) R\ ,
\end{equation}
where $R$ is the ten-dimensional Ricci scalar.  Couplings of this type
for the tachyon of the bosonic string were calculated using sigma model
techniques in \cite{Tseytlin:2000mt,Andreev:2003hn}.\footnote{Early computations
of the tachyon potential can be found in \cite{Bardakci:1974gv,Bardakci:1974vs,
Bardakci:1975ux,Bardakci:1977an}.}  The lack of a term surviving the $R_{ads}\to\infty$ limit is still a mystery.  
We note, however, that Yang and Zwiebach \cite{Yang:2005rx}\ claim that
the tachyon potential vanishes both in the perturbative closed string vacuum and
after the tachyon has condensed.

We find this conjectured coupling interesting, as similar couplings appear
between massless open strings on unstable D-branes and the
open string tachyons mediating the decay of these branes.
The tree-level effective action
for massless and tachyon modes in this system is well described by
the Born-Infeld-like Lagrangian:\cite{Sen:1999md,Harvey:2000na,Kutasov:2000qp,
Cornalba:2000ad,Okuyama:2000ch,Witten:1992qy}
\begin{equation}
  S = \int d^d x V(T) \sqrt{\det\left(\eta_{\mu\nu} - 2\pi \alpha' F_{\mu\nu}\right)
    }
\end{equation}
Here $V(T)$ vanishes as the tachyon condenses and the branes disappear,
indicating that the open string degrees of freedom disappear from the
spectrum \cite{Sen:1999md,Sen:1998sm,Sen:1999xm,Sen:1999nx,
Berkovits:2000hf,Yi:1999hd,Bergman:2000xf,Gopakumar:2000rw,
David:2000uv,Taylor:2000ku,Sen:2000hx,Kleban:2000pf}.

A similar story arises when one studies the worldsheet theory of open strings in the
presence of a tachyon background, discussed in
\cite{Harvey:2000na,Kutasov:2000qp}.  In that work, the
tachyon condensate led to a potential on the boundary of the worldsheet,
making it energetically unfavorable for the boundary of the string to live where
the open string tachyon condenses.  Adams {\it et. al.} \cite{Adams:2005rb}\
and McGreevy and Silverstein \cite{McGreevy:2005ci}.
have argued that a similar phenomenon occurs for localized tachyonic
winding states in closed string theory.\footnote{A similar proposal has been made 
for the bosonic string tachyon by Yang and Zwiebach,
based on calculations in closed string field theory \cite{Yang:2005rx}.}
In particular, the tachyon condensate
induces a potential on the worldsheet, suppressing fluctuations of the string into
regions with nonzero condensate.  Worldsheet correlation functions have support
away from this condensate, indicating that closed string amplitudes
vanish because the closed strings cease to become dynamical when
the tachyon condenses.  Indeed, if the condensate of winding tachyons describes
the Euclidean black hole, the spacetime ends at the horizon, which we wish to
identify as the location of or boundary of the tachyon condensate.
 
If the proposed action (\ref{eq:tachact}) is correct, and if $V(T)$ vanishes
at the minimum of the tachyon potential, a possible interpretation of this
state is that the closed string degrees of freedom disappear, along the
lines of the open string scenario.  This is consistent with
indications based on the worldsheet theory \cite{Kutasov:1992pf,Hsu:1992cm,
Kachru:1999ed,Harvey:2000na}.  

A serious argument for this gravitational ``nothing state'' as the endpoint of
tachyon condensation
will require more than the existence of (\ref{eq:tachact}).  
A fuller test of this proposal would be to compute the coupling of the tachyon to
the full effective action for the closed string graviton and
other massless closed string modes.  If this
action is, schematically
\begin{equation}
  S = \int d^d x \sqrt{g} V(\phi) F(R)
\end{equation}
then we have a better indication that the gravitational dynamics truly decouples
at the endpoint $V=0$ of the condensation process.
It would be very interesting to check this coupling from either the worldsheet
point of view (perhaps computing multipoint correlators along the lines of \cite{McGreevy:2005ci})
or via string field theory.\footnote{Some preliminary calculations for localized tachyons in closed
string field theory appear in \cite{Okawa:2004rh}.}

\section{Conclusions}

In the first part of this work, we demonstrated that the dynamics of the
thermal scalar for finite-temperature string theory was dual
to the dynamics of long strings, and we used this duality to
develop a picture of thermally excited strings as random
walks describing the configurations of these strings.  Following
this, we made some observations regarding the endpoint of 
the condensation of the thermal scalar in AdS spacetimes where
string thermodynamics is under some control.  In particular, we have
argued that the thermal scalar condensate is somehow dual
to the AdS-Schwarzchild black hole.  It would be interesting to
understand this relationship better.  The relationship between the thermal
scalar and long strings might give a better microscopic picture of the
origins of black hole entropy.  It would also be interesting to pursue further
the suggested decoupling of closed string dynamics when the
Hagedorn tachyon condenses.

\section{Acknowledgements}

We are especially grateful to Ofer Aharony, John McGreevy, Howard Schnitzer,
and Eva Silverstein for comments on a draft of this manuscript, and for very useful
discussions and comments. We would also like to thank Allan Adams,  Bulbul Chakraborty, 
Shiraz Minwalla, and Rajesh Ravindran for helpful conversations on these
and related subjects.  A.L. would like to thank the Perimeter Institute for
Theoretical Physics and the Fields Institute for Research in the Mathematical
Sciences for their hospitality while part of this
work was being carried out.  This work was supported in part by NSF grant PHY-0331516,
DOE grant DE-FG02-92ER40706, and an Outstanding Junior Investigator award. In addition,
M.K. was also supported by NSF grant PHY-0243680. Any opinions, 
findings, and conclusions or recommendations expressed in this material are those of the 
authors and do not necessarily reflect the views of the National Science Foundation.

\appendix

\section{An identity for the one- and two-string partition functions}

Let us study the two-string partition function.  Choose a basis of
energy eigenstates $\ket{k}$ for a single string such that the Hamiltonian $H$ has
energy eigenvalues $E_k$.  The basis elements
will include different string modes.  
The partition funnction for a single string is
\begin{equation}
  Z_{\beta,1} = \sum_k e^{-\beta E_k}
\end{equation}

A basis of states for two bosonic strings with Hamiltonian $H = H_A + H_B$
(where $H_A,H_B$ have identical spectra) is:
\begin{eqnarray}
  \ket{k,l} &=& \frac{1}{\sqrt{2}} \left(\ket{k}_A\ket{l}_B + \ket{l}_B\ket{k}_A\right)\
  \ \ \ \ \ \ \ k \neq l \\
  \ket{k,k} &=& \ket{k}_1\ket{k}_2
\end{eqnarray}
The partition function for two strings is 
\begin{eqnarray}
  Z_{\beta,2} &=& \sum_{k>l} e^{-\beta(E_k + E_l)}
       + \sum_k e^{-2\beta E_k}\\
       &=& \half\left(Z_{\beta,1}^2 - \sum_k e^{-2\beta E_k}\right) + \sum_k e^{-2\beta E_k}\\
       &=& \half\left(Z_{\beta,1}^2 + \sum_k e^{-2\beta E_k}\right)
\end{eqnarray} 
Therefore
\begin{equation}
  Z_{\beta,2} - \half Z_{\beta,1}^2 = \half Z_{2\beta,1}^2
\end{equation}
Inspection of (\ref{eq:oneparticlepf},\ref{eq:onepmodular}) reveals that 
$Z_{2\beta} = 2 Z_{[0,2]}$, proving the last line of (\ref{eq:freetopart}).


\begin{thebibliography}{99}

%\cite{Polchinski:1985zf}
\bibitem{Polchinski:1985zf}
J.~Polchinski,
``Evaluation Of The One Loop String Path Integral,''
Commun.\ Math.\ Phys.\  {\bf 104}, 37 (1986).
%%CITATION = CMPHA,104,37;%%

%\cite{Sathiapalan:1986db}
\bibitem{Sathiapalan:1986db}
  B.~Sathiapalan,
  ``Vortices On The String World Sheet And Constraints On Toral
  Compactification,''
  Phys.\ Rev.\ D {\bf 35}, 3277 (1987).
  %%CITATION = PHRVA,D35,3277;%%

%\cite{Kogan:1987jd}
\bibitem{Kogan:1987jd}
  Y.~I.~Kogan,
  ``Vortices On The World Sheet And String's Critical Dynamics,''
  JETP Lett.\  {\bf 45}, 709 (1987)
  [Pisma Zh.\ Eksp.\ Teor.\ Fiz.\  {\bf 45}, 556 (1987)].
  %%CITATION = JTPLA,45,709;%%

%\cite{Atick:1988si}
\bibitem{Atick:1988si}
J.~J.~Atick and E.~Witten,
``The Hagedorn Transition And The Number Of Degrees Of Freedom Of String
Theory,''
Nucl.\ Phys.\ B {\bf 310}, 291 (1988).
%%CITATION = NUPHA,B310,291;%%

%\cite{Brandenberger:1988aj}
\bibitem{Brandenberger:1988aj}
R.~H.~Brandenberger and C.~Vafa,
``Superstrings In The Early Universe,''
Nucl.\ Phys.\ B {\bf 316}, 391 (1989).
%%CITATION = NUPHA,B316,391;%%

%\cite{Horowitz:1997jc}
\bibitem{Horowitz:1997jc}
G.~T.~Horowitz and J.~Polchinski,
``Self gravitating fundamental strings,''
Phys.\ Rev.\ D {\bf 57}, 2557 (1998)
[arXiv:hep-th/9707170].
%%CITATION = HEP-TH 9707170;%%

%\cite{Mitchell:1987hr}
\bibitem{Mitchell:1987hr}
D.~Mitchell and N.~Turok,
``Statistical Mechanics Of Cosmic Strings,''
Phys.\ Rev.\ Lett.\  {\bf 58}, 1577 (1987).
%%CITATION = PRLTA,58,1577;%%

%\cite{Mitchell:1987th}
\bibitem{Mitchell:1987th}
D.~Mitchell and N.~Turok,
``Statistical Properties Of Cosmic Strings,''
Nucl.\ Phys.\ B {\bf 294}, 1138 (1987).
%%CITATION = NUPHA,B294,1138;%%

%\cite{Bowick:1989us}
\bibitem{Bowick:1989us}
M.~J.~Bowick and S.~B.~Giddings,
``High Temperature Strings,''
Nucl.\ Phys.\ B {\bf 325}, 631 (1989).
%%CITATION = NUPHA,B325,631;%%

%\cite{Barbon:2004dd}
\bibitem{Barbon:2004dd}
  J.~L.~F.~Barbon and E.~Rabinovici,
  ``Touring the Hagedorn ridge,''
  arXiv:hep-th/0407236.
  %%CITATION = HEP-TH 0407236;%%

%\cite{Maldacena:1997re}
\bibitem{Maldacena:1997re}
  J.~M.~Maldacena,
  %``The large N limit of superconformal field theories and supergravity,''
  Adv.\ Theor.\ Math.\ Phys.\  {\bf 2}, 231 (1998)
  [Int.\ J.\ Theor.\ Phys.\  {\bf 38}, 1113 (1999)]
  [arXiv:hep-th/9711200].
  %%CITATION = HEP-TH 9711200;%%

%\cite{Sundborg:1999ue}
\bibitem{Sundborg:1999ue}
B.~Sundborg,
``The Hagedorn transition, deconfinement and N = 4 SYM theory,''
Nucl.\ Phys.\ B {\bf 573}, 349 (2000)
[arXiv:hep-th/9908001].
%%CITATION = HEP-TH 9908001;%%

%\cite{Haggi-Mani:2000ru}
\bibitem{Haggi-Mani:2000ru}
P.~Haggi-Mani and B.~Sundborg,
``Free large N supersymmetric Yang-Mills theory as a string theory,''
JHEP {\bf 0004}, 031 (2000)
[arXiv:hep-th/0002189].
%%CITATION = HEP-TH 0002189;%%

%\cite{Aharony:2003sx}
\bibitem{Aharony:2003sx}
O.~Aharony, J.~Marsano, S.~Minwalla, K.~Papadodimas and M.~Van  
Raamsdonk,
``The Hagedorn/deconfinement phase transition in weakly coupled large N  
gauge theories,'' arXiv:hep-th/0310285.
%%CITATION = HEP-TH 0310285;%%

  %\cite{Schnitzer:2004qt}
\bibitem{Schnitzer:2004qt}
  H.~J.~Schnitzer,
  ``Confinement / deconfinement transition of large N gauge theories with  N(f)
  fundamentals: N(f)/N finite,''
  Nucl.\ Phys.\ B {\bf 695}, 267 (2004)
  [arXiv:hep-th/0402219].
  %%CITATION = HEP-TH 0402219;%%

%\cite{Spradlin:2004pp}
\bibitem{Spradlin:2004pp}
M.~Spradlin and A.~Volovich,
``A pendant for Polya: The one-loop partition function of N = 4 SYM on  
R x S(3),''  arXiv:hep-th/0408178.
%%CITATION = HEP-TH 0408178;%%

%\cite{Liu:2004vy}
\bibitem{Liu:2004vy}
H.~Liu,
``Fine structure of Hagedorn transitions,''
arXiv:hep-th/0408001.
%%CITATION = HEP-TH 0408001;%%

%\cite{Aharony:2005bq}
\bibitem{Aharony:2005bq}
  O.~Aharony, J.~Marsano, S.~Minwalla, K.~Papadodimas and M.~Van Raamsdonk,
  ``A first order deconfinement transition in large N Yang-Mills theory on a
  small S**3,''
  arXiv:hep-th/0502149.
  %%CITATION = HEP-TH 0502149;%%
   
%\cite{Alvarez-Gaume:2005fv}
\bibitem{Alvarez-Gaume:2005fv}
  L.~Alvarez-Gaume, C.~Gomez, H.~Liu and S.~Wadia,
  ``Finite temperature effective action, AdS(5) black holes, and 1/N
  expansion,''
  arXiv:hep-th/0502227.
  %%CITATION = HEP-TH 0502227;%%
  


%\cite{Gomez-Reino:2005bq}
\bibitem{Gomez-Reino:2005bq}
  M.~Gomez-Reino, S.~G.~Naculich and H.~J.~Schnitzer,
  ``More pendants for Polya: Two loops in the SU(2) sector,''
  arXiv:hep-th/0504222.
  %%CITATION = HEP-TH 0504222;%%
  
  
%\cite{Barbon:2001di}
\bibitem{Barbon:2001di}
  J.~L.~F.~Barbon and E.~Rabinovici,
  ``Closed-string tachyons and the Hagedorn transition in AdS space,''
  JHEP {\bf 0203}, 057 (2002)
  [arXiv:hep-th/0112173].
  %%CITATION = HEP-TH 0112173;%%
  
%\cite{Barbon:2002nw}
\bibitem{Barbon:2002nw}
  J.~L.~F.~Barbon and E.~Rabinovici,
  ``Remarks on black hole instabilities and closed string tachyons,''
  Found.\ Phys.\  {\bf 33}, 145 (2003)
  [arXiv:hep-th/0211212].
  %%CITATION = HEP-TH 0211212;%%
  
%\cite{Adams:2005rb}
\bibitem{Adams:2005rb}
  A.~Adams, X.~Liu, J.~McGreevy, A.~Saltman and E.~Silverstein,
  ``Things fall apart: Topology change from winding tachyons,''
  arXiv:hep-th/0502021.
  %%CITATION = HEP-TH 0502021;%%
  
%\cite{Sen:1998sm}
\bibitem{Sen:1998sm}
  A.~Sen,
  ``Tachyon condensation on the brane antibrane system,''
  JHEP {\bf 9808}, 012 (1998)
  [arXiv:hep-th/9805170].
  %%CITATION = HEP-TH 9805170;%%
  
%\cite{Sen:1999md}
\bibitem{Sen:1999md}
  A.~Sen,
  ``Supersymmetric world-volume action for non-BPS D-branes,''
  JHEP {\bf 9910}, 008 (1999)
  [arXiv:hep-th/9909062].
  %%CITATION = HEP-TH 9909062;%%
  
 %\cite{1999xm}
\bibitem{Sen:1999xm}
  A.~Sen,
  ``Universality of the tachyon potential,''
  JHEP {\bf 9912}, 027 (1999)
  [arXiv:hep-th/9911116].
  %%CITATION = HEP-TH 9911116;%%

%\cite{Sen:1999nx}
\bibitem{Sen:1999nx}
  A.~Sen and B.~Zwiebach,
  ``Tachyon condensation in string field theory,''
  JHEP {\bf 0003}, 002 (2000)
  [arXiv:hep-th/9912249].
  %%CITATION = HEP-TH 9912249;%%

%\cite{Berkovits:2000hf}
\bibitem{Berkovits:2000hf}
  N.~Berkovits, A.~Sen and B.~Zwiebach,
  ``Tachyon condensation in superstring field theory,''
  Nucl.\ Phys.\ B {\bf 587}, 147 (2000)
  [arXiv:hep-th/0002211].
  %%CITATION = HEP-TH 0002211;%%

%\cite{Yi:1999hd}
\bibitem{Yi:1999hd}
  P.~Yi,
  ``Membranes from five-branes and fundamental strings from Dp branes,''
  Nucl.\ Phys.\ B {\bf 550}, 214 (1999)
  [arXiv:hep-th/9901159].
  %%CITATION = HEP-TH 9901159;%%

%\cite{Bergman:2000xf}
\bibitem{Bergman:2000xf}
  O.~Bergman, K.~Hori and P.~Yi,
  %``Confinement on the brane,''
  Nucl.\ Phys.\ B {\bf 580}, 289 (2000)
  [arXiv:hep-th/0002223].
  %%CITATION = HEP-TH 0002223;%%

%\cite{Gopakumar:2000rw}
\bibitem{Gopakumar:2000rw}
  R.~Gopakumar, S.~Minwalla and A.~Strominger,
  ``Symmetry restoration and tachyon condensation in open string theory,''
  JHEP {\bf 0104}, 018 (2001)
  [arXiv:hep-th/0007226].
  %%CITATION = HEP-TH 0007226;%%

%\cite{David:2000uv}
\bibitem{David:2000uv}
  J.~R.~David,
  ``U(1) gauge invariance from open string field theory,''
  JHEP {\bf 0010}, 017 (2000)
  [arXiv:hep-th/0005085].
  %%CITATION = HEP-TH 0005085;%%

%\cite{Taylor:2000ku}
\bibitem{Taylor:2000ku}
  W.~Taylor,
  ``Mass generation from tachyon condensation for vector fields on  D-branes,''
  JHEP {\bf 0008}, 038 (2000)
  [arXiv:hep-th/0008033].
  %%CITATION = HEP-TH 0008033;%%

%\cite{Sen:2000hx}
\bibitem{Sen:2000hx}
  A.~Sen and B.~Zwiebach,
  ``Large marginal deformations in string field theory,''
  JHEP {\bf 0010}, 009 (2000)
  [arXiv:hep-th/0007153].
  %%CITATION = HEP-TH 0007153;%%

%\cite{Kleban:2000pf}
\bibitem{Kleban:2000pf}
  M.~Kleban, A.~E.~Lawrence and S.~H.~Shenker,
  ``Closed strings from nothing,''
  Phys.\ Rev.\ D {\bf 64}, 066002 (2001)
  [arXiv:hep-th/0012081].
  %%CITATION = HEP-TH 0012081;%%
  
%\cite{Horowitz:2005vp}
\bibitem{Horowitz:2005vp}
  G.~T.~Horowitz,
  ``Tachyon condensation and black strings,''
  arXiv:hep-th/0506166.
  %%CITATION = HEP-TH 0506166;%%
  
%\cite{McGreevy:2005ci}
\bibitem{McGreevy:2005ci}
  J.~McGreevy and E.~Silverstein,
  ``The tachyon at the end of the universe,''
  arXiv:hep-th/0506130.
  %%CITATION = HEP-TH 0506130;%%

%\cite{Manes:2004nd}
\bibitem{Manes:2004nd}
  J.~L.~Manes,
  ``Portrait of the string as a random walk,''
  JHEP {\bf 0503}, 070 (2005)
  [arXiv:hep-th/0412104].
  %%CITATION = HEP-TH 0412104;%%

  %\cite{McClain:1986id}
\bibitem{McClain:1986id}
  B.~McClain and B.~D.~B.~Roth,
  ``Modular Invariance For Interacting Bosonic Strings At Finite Temperature,''
  Commun.\ Math.\ Phys.\  {\bf 111}, 539 (1987).
  %%CITATION = CMPHA,111,539;%%

   %\cite{Freund:1980xh}
\bibitem{Freund:1980xh}
  P.~G.~O.~Freund and M.~A.~Rubin,
  ``Dynamics Of Dimensional Reduction,''
  Phys.\ Lett.\ B {\bf 97}, 233 (1980).
  %%CITATION = PHLTA,B97,233;%%
  
  %\cite{Monthus:1996}
\bibitem{Monthus:1996}
C.~Monthus and C.~Texier, "Random Walk on the Bethe Lattice and
Hyperbolic Brownian Motion," J.\ Phys.\ A: Math.\ Gen.\ {\bf 29}, 2399  
(1996).

  
 %\cite{Dowker:1975xj}
\bibitem{Dowker:1975xj}
  J.~S.~Dowker and R.~Critchley,
  ``Scalar Effective Lagrangian In De Sitter Space,''
  Phys.\ Rev.\ D {\bf 13}, 224 (1976).
  %%CITATION = PHRVA,D13,224;%%
  
%\cite{Dowker:1975tf}
\bibitem{Dowker:1975tf}
  J.~S.~Dowker and R.~Critchley,
  ``Effective Lagrangian And Energy Momentum Tensor In De Sitter Space,''
  Phys.\ Rev.\ D {\bf 13}, 3224 (1976).
  %%CITATION = PHRVA,D13,3224;%%


%\cite{Breitenlohner:1982bm}
\bibitem{Breitenlohner:1982bm}
  P.~Breitenlohner and D.~Z.~Freedman,
  ``Positive Energy In Anti-De Sitter Backgrounds And Gauged Extended
  Supergravity,''
  Phys.\ Lett.\ B {\bf 115}, 197 (1982).
  %%CITATION = PHLTA,B115,197;%%

%\cite{Breitenlohner:1982jf}
\bibitem{Breitenlohner:1982jf}
  P.~Breitenlohner and D.~Z.~Freedman,
  ``Stability In Gauged Extended Supergravity,''
  Annals Phys.\  {\bf 144}, 249 (1982).
  %%CITATION = APNYA,144,249;%%

%\cite{Adams:2001sv}
\bibitem{Adams:2001sv}
  A.~Adams, J.~Polchinski and E.~Silverstein,
  ``Don't panic! Closed string tachyons in ALE space-times,''
  JHEP {\bf 0110}, 029 (2001)
  [arXiv:hep-th/0108075].
  %%CITATION = HEP-TH 0108075;%%

%Igor's type 0 papers
%\cite{Klebanov:1998yy}
\bibitem{Klebanov:1998yy}
  I.~R.~Klebanov and A.~A.~Tseytlin,
  ``D-branes and dual gauge theories in type 0 strings,''
  Nucl.\ Phys.\ B {\bf 546}, 155 (1999)
  [arXiv:hep-th/9811035].
  %%CITATION = HEP-TH 9811035;%%

%\cite{Klebanov:1999ch}
\bibitem{Klebanov:1999ch}
  I.~R.~Klebanov and A.~A.~Tseytlin,
  ``A non-supersymmetric large N CFT from type 0 string theory,''
  JHEP {\bf 9903}, 015 (1999)
  [arXiv:hep-th/9901101].
  %%CITATION = HEP-TH 9901101;%%


%\cite{Klebanov:1999um}
\bibitem{Klebanov:1999um}
  I.~R.~Klebanov,
  ``Tachyon stabilization in the AdS/CFT correspondence,''
  Phys.\ Lett.\ B {\bf 466}, 166 (1999)
  [arXiv:hep-th/9906220].
  %%CITATION = HEP-TH 9906220;%%
  
  %\cite{Tseytlin:1999ii}
\bibitem{Tseytlin:1999ii}
  A.~A.~Tseytlin and K.~Zarembo,
  ``Effective potential in non-supersymmetric SU(N) x SU(N) gauge theory  and
  interactions of type 0 D3-branes,''
  Phys.\ Lett.\ B {\bf 457}, 77 (1999)
  [arXiv:hep-th/9902095].
  %%CITATION = HEP-TH 9902095;%%
  
%\cite{Adams:2001jb}
\bibitem{Adams:2001jb}
  A.~Adams and E.~Silverstein,
  ``Closed string tachyons, AdS/CFT, and large N QCD,''
  Phys.\ Rev.\ D {\bf 64}, 086001 (2001)
  [arXiv:hep-th/0103220].
  %%CITATION = HEP-TH 0103220;%%

%\cite{Polyakov:1978vu}
\bibitem{Polyakov:1978vu}
  A.~M.~Polyakov,
  ``Thermal Properties Of Gauge Fields And Quark Liberation,''
  Phys.\ Lett.\ B {\bf 72}, 477 (1978).
  %%CITATION = PHLTA,B72,477;%%

%\cite{Susskind:1979up}
\bibitem{Susskind:1979up}
  L.~Susskind,
  ``Lattice Models Of Quark Confinement At High Temperature,''
  Phys.\ Rev.\ D {\bf 20}, 2610 (1979).
  %%CITATION = PHRVA,D20,2610;%%

%\cite{Witten:1998zw}
\bibitem{Witten:1998zw}
  E.~Witten,
  ``Anti-de Sitter space, thermal phase transition, and confinement in  gauge
  theories,''
  Adv.\ Theor.\ Math.\ Phys.\  {\bf 2}, 505 (1998)
  [arXiv:hep-th/9803131].
  %%CITATION = HEP-TH 9803131;%%

%\cite{Witten:1998qj}
\bibitem{Witten:1998qj}
   E.~Witten,
   ``Anti-de Sitter space and holography,''
   Adv.\ Theor.\ Math.\ Phys.\  {\bf 2}, 253 (1998)
   [arXiv:hep-th/9802150].
   %%CITATION = HEP-TH 9802150;%%

%\cite{Maldacena:1998im}
\bibitem{Maldacena:1998im}
  J.~M.~Maldacena,
  ``Wilson loops in large N field theories,''
  Phys.\ Rev.\ Lett.\  {\bf 80}, 4859 (1998)
  [arXiv:hep-th/9803002].
  %%CITATION = HEP-TH 9803002;%%

%\cite{Hawking:1982dh}
\bibitem{Hawking:1982dh}
  S.~W.~Hawking and D.~N.~Page,
  ``Thermodynamics Of Black Holes In Anti-De Sitter Space,''
  Commun.\ Math.\ Phys.\  {\bf 87}, 577 (1983).
  %%CITATION = CMPHA,87,577;%%

%\cite{Horowitz:1996nw}
\bibitem{Horowitz:1996nw}
  G.~T.~Horowitz and J.~Polchinski,
  ``A correspondence principle for black holes and strings,''
  Phys.\ Rev.\ D {\bf 55}, 6189 (1997)
  [arXiv:hep-th/9612146].
  %%CITATION = HEP-TH 9612146;%%
  
%\cite{Rey:1998ik}
\bibitem{Rey:1998ik}
  S.~J.~Rey and J.~T.~Yee,
  ``Macroscopic strings as heavy quarks in large N gauge theory and  anti-de
  Sitter supergravity,''
  Eur.\ Phys.\ J.\ C {\bf 22}, 379 (2001)
  [arXiv:hep-th/9803001].
  %%CITATION = HEP-TH 9803001;%%

%\cite{Rey:1998bq}
\bibitem{Rey:1998bq}
  S.~J.~Rey, S.~Theisen and J.~T.~Yee,
  ``Wilson-Polyakov loop at finite temperature in large N gauge theory and
  anti-de Sitter supergravity,''
  Nucl.\ Phys.\ B {\bf 527}, 171 (1998)
  [arXiv:hep-th/9803135].
  %%CITATION = HEP-TH 9803135;%%
  
 %\cite{Brandhuber:1998bs}
\bibitem{Brandhuber:1998bs}
  A.~Brandhuber, N.~Itzhaki, J.~Sonnenschein and S.~Yankielowicz,
  ``Wilson loops in the large N limit at finite temperature,''
  Phys.\ Lett.\ B {\bf 434}, 36 (1998)
  [arXiv:hep-th/9803137].
  %%CITATION = HEP-TH 9803137;%%
  
%\cite{Brandhuber:1998er}
\bibitem{Brandhuber:1998er}
  A.~Brandhuber, N.~Itzhaki, J.~Sonnenschein and S.~Yankielowicz,
  ``Wilson loops, confinement, and phase transitions in large N gauge  theories
  from supergravity,''
  JHEP {\bf 9806}, 001 (1998)
  [arXiv:hep-th/9803263].
  %%CITATION = HEP-TH 9803263;%%

%\cite{Ebrahim:2005uz}
\bibitem{Ebrahim:2005uz}
  H.~Ebrahim and A.~E.~Mosaffa,
  ``Semiclassical string solutions on 1/2 BPS geometries,''
  JHEP {\bf 0501}, 050 (2005)
  [arXiv:hep-th/0501072].
  %%CITATION = HEP-TH 0501072;%%
  
%\cite{Drukker:1999zq}
\bibitem{Drukker:1999zq}
  N.~Drukker, D.~J.~Gross and H.~Ooguri,
  ``Wilson loops and minimal surfaces,''
  Phys.\ Rev.\ D {\bf 60}, 125006 (1999)
  [arXiv:hep-th/9904191].
  %%CITATION = HEP-TH 9904191;%%
  
  %\cite{Kosterlitz:1973xp}
\bibitem{Kosterlitz:1973xp}
  J.~M.~Kosterlitz and D.~J.~Thouless,
  ``Ordering, Metastability And Phase Transitions In Two-Dimensional
  Systems,''
  J.\ Phys.\  C {\bf 6} (1973) 1181.
  %%CITATION = JPCBA,C6,1181;%%

%\cite{Kosterlitz:1974sm}
\bibitem{Kosterlitz:1974sm}
  J.~M.~Kosterlitz,
  ``The Critical Properties Of The Two-Dimensional Xy Model,''
  J.\ Phys.\ C {\bf 7}, 1046 (1974).
  %%CITATION = JPCBA,7,1046;%%

%\cite{Harvey:2000na}
\bibitem{Harvey:2000na}
  J.~A.~Harvey, D.~Kutasov and E.~J.~Martinec,
  ``On the relevance of tachyons,''
  arXiv:hep-th/0003101.
  %%CITATION = HEP-TH 0003101;%%
  
  %\cite{Strominger:1996sh}
\bibitem{Strominger:1996sh}
  A.~Strominger and C.~Vafa,
  ``Microscopic Origin of the Bekenstein-Hawking Entropy,''
  Phys.\ Lett.\ B {\bf 379}, 99 (1996)
  [arXiv:hep-th/9601029].
  %%CITATION = HEP-TH 9601029;%%

%\cite{Hawking:1994ii}
\bibitem{Hawking:1994ii}
  S.~W.~Hawking, G.~T.~Horowitz and S.~F.~Ross,
  ``Entropy, Area, and black hole pairs,''
  Phys.\ Rev.\ D {\bf 51}, 4302 (1995)
  [arXiv:gr-qc/9409013].
  %%CITATION = GR-QC 9409013;%%
  
%\cite{Dabholkar:2001if}
\bibitem{Dabholkar:2001if}
  A.~Dabholkar,
  ``Tachyon condensation and black hole entropy,''
  Phys.\ Rev.\ Lett.\  {\bf 88}, 091301 (2002)
  [arXiv:hep-th/0111004].
  %%CITATION = HEP-TH 0111004;%%

%\cite{Susskind:1994sm}
\bibitem{Susskind:1994sm}
  L.~Susskind and J.~Uglum,
  ``Black hole entropy in canonical quantum gravity and superstring theory,''
  Phys.\ Rev.\ D {\bf 50}, 2700 (1994)
  [arXiv:hep-th/9401070].
  %%CITATION = HEP-TH 9401070;%%

%\cite{Tseytlin:2000mt}
\bibitem{Tseytlin:2000mt}
  A.~A.~Tseytlin,
  ``Sigma model approach to string theory effective actions with tachyons,''
  J.\ Math.\ Phys.\  {\bf 42}, 2854 (2001)
  [arXiv:hep-th/0011033].
  %%CITATION = HEP-TH 0011033;%%

%\cite{Andreev:2003hn}
\bibitem{Andreev:2003hn}
  O.~Andreev,
  ``Comments on tachyon potentials in closed and open-closed string
  theories,''
  Nucl.\ Phys.\ B {\bf 680}, 3 (2004)
  [arXiv:hep-th/0308123].
  %%CITATION = HEP-TH 0308123;%%

%\cite{Bardakci:1974gv}
\bibitem{Bardakci:1974gv}
  K.~Bardakci,
  ``Dual Models And Spontaneous Symmetry Breaking,''
  Nucl.\ Phys.\ B {\bf 68}, 331 (1974).
  %%CITATION = NUPHA,B68,331;%%

%\cite{Bardakci:1974vs}
\bibitem{Bardakci:1974vs}
  K.~Bardakci and M.~B.~Halpern,
  ``Explicit Spontaneous Breakdown In A Dual Model,''
  Phys.\ Rev.\ D {\bf 10}, 4230 (1974).
  %%CITATION = PHRVA,D10,4230;%%

%\cite{Bardakci:1975ux}
\bibitem{Bardakci:1975ux}
  K.~Bardakci and M.~B.~Halpern,
  ``Explicit Spontaneous Breakdown In A Dual Model. 2. N Point Functions,''
  Nucl.\ Phys.\ B {\bf 96}, 285 (1975).
  %%CITATION = NUPHA,B96,285;%%

%\cite{Bardakci:1977an}
\bibitem{Bardakci:1977an}
  K.~Bardakci,
  ``Spontaneous Symmetry Breakdown In The Standard Dual String Model,''
  Nucl.\ Phys.\ B {\bf 133}, 297 (1978).
  %%CITATION = NUPHA,B133,297;%%


%\cite{Yang:2005rx}
\bibitem{Yang:2005rx}
  H.~Yang and B.~Zwiebach,
  ``A closed string tachyon vacuum?,''
  arXiv:hep-th/0506077.
  %%CITATION = HEP-TH 0506077;%%


%\cite{Kutasov:2000qp}
\bibitem{Kutasov:2000qp}
  D.~Kutasov, M.~Marino and G.~W.~Moore,
  ``Some exact results on tachyon condensation in string field theory,''
  JHEP {\bf 0010}, 045 (2000)
  [arXiv:hep-th/0009148].
  %%CITATION = HEP-TH 0009148;%%

%\cite{Cornalba:2000ad}
\bibitem{Cornalba:2000ad}
  L.~Cornalba,
  ``Tachyon condensation in large magnetic fields with background  independent
  string field theory,''
  Phys.\ Lett.\ B {\bf 504}, 55 (2001)
  [arXiv:hep-th/0010021].
  %%CITATION = HEP-TH 0010021;%%

%\cite{Okuyama:2000ch}
\bibitem{Okuyama:2000ch}
  K.~Okuyama,
  ``Noncommutative tachyon from background independent open string field
  theory,''
  Phys.\ Lett.\ B {\bf 499}, 167 (2001)
  [arXiv:hep-th/0010028].
  %%CITATION = HEP-TH 0010028;%%

%\cite{Witten:1992qy}
\bibitem{Witten:1992qy}
  E.~Witten,
  ``On background independent open string field theory,''
  Phys.\ Rev.\ D {\bf 46}, 5467 (1992)
  [arXiv:hep-th/9208027].
  %%CITATION = HEP-TH 9208027;%%

%\cite{Kachru:1999ed}
\bibitem{Kachru:1999ed}
  S.~Kachru, J.~Kumar and E.~Silverstein,
  ``Orientifolds, RG flows, and closed string tachyons,''
  Class.\ Quant.\ Grav.\  {\bf 17}, 1139 (2000)
  [arXiv:hep-th/9907038].
  %%CITATION = HEP-TH 9907038;%%

%\cite{Kutasov:1992pf}
\bibitem{Kutasov:1992pf}
  D.~Kutasov,
  ``Irreversibility of the renormalization group flow in two-dimensional
  quantum gravity,''
  Mod.\ Phys.\ Lett.\ A {\bf 7}, 2943 (1992)
  [arXiv:hep-th/9207064].
  %%CITATION = HEP-TH 9207064;%%

%\cite{Hsu:1992cm}
\bibitem{Hsu:1992cm}
  E.~Hsu and D.~Kutasov,
  ``The Gravitational Sine-Gordon model,''
  Nucl.\ Phys.\ B {\bf 396}, 693 (1993)
  [arXiv:hep-th/9212023].
  %%CITATION = HEP-TH 9212023;%%

%\cite{Okawa:2004rh}
\bibitem{Okawa:2004rh}
  Y.~Okawa and B.~Zwiebach,
   `Twisted tachyon condensation in closed string field theory,''
  JHEP {\bf 0403}, 056 (2004)
  [arXiv:hep-th/0403051].
  %%CITATION = HEP-TH 0403051;%%

%LAST REF

  

\end{thebibliography}
\end{document}